\documentclass[]{spie}  

\usepackage[dvips]{graphicx}

\def\micron{{$\mu$m}}

\def\degree{{${}^\circ$}}

\def\flux{counts/s/cm${}^2$/keV}

\begin{document}

\thispagestyle{empty}
\renewcommand{\thefootnote}{\fnsymbol{footnote}}

\begin{flushright}
{\small
SLAC--PUB--10539\\
July, 2004\\}
\end{flushright}

\vspace{.8cm}

\begin{center}
{\bf\large   
Gamma-ray Polarimetry with Compton Telescope
\footnote{Work supported by
Department of Energy contract  DE--AC03--76SF00515, Grantin-Aid by Ministry of Education, Culture, Sports, Science and Technology of Japan (12554006, 13304014),
and ``Ground-based Research Announcement for Space Utilization'' promoted by Japan Space Forum.}}

\vspace{1cm}

Hiroyasu Tajima\supit{a}, 
Grzegorz~Madejski\supit{a},
Takefumi Mitani\supit{b,c}, 
Takaaki Tanaka\supit{b,c}, 
Hidehito~Nakamura\supit{d}, 
Kazuhiro~Nakazawa\supit{b},
Tadayuki Takahashi\supit{b,c}, 
Yasushi Fukazawa\supit{e}, 
Tuneyoshi~Kamae\supit{a}, 
Motohide~Kokubun\supit{c},
Daniel Marlow\supit{f},
Masaharu Nomachi\supit{d}
and Eduardo~do~Couto~e~Silva\supit{a}
\skiplinehalf
\supit{a} Stanford Linear Accelerator Center, Stanford, CA 94309-4349, USA\\
\supit{b} Institute of Space and Astronautical Science, Sagamihara, Kanagawa 229-8510, Japan\\
\supit{c} Department of Physics, University of Tokyo, Bunkyo-ku, Tokyo 113-0033, Japan\\
\supit{d} Department of Physics, Osaka University, Toyonaka, Osaka 560-0043, Japan\\
\supit{e} Department of Physics, Hiroshima University, Higashi-Hiroshima 739-8526, Japan\\
\supit{f} Department of Physics, Priceton University, Princeton, NJ 08544\\

\end{center}

\vfill

\begin{center}
{\bf\large   
Abstract }
\end{center}

\begin{quote}
Compton telescope is a promising technology to achieve very high sensitivity in the soft gamma-ray band (0.1--10 MeV) by utilizing Compton kinematics. Compton kinematics also enables polarization measurement which will open new windows to study gamma-ray production mechanism in the universe.
CdTe and Si semiconductor technologies are key technologies to realize the Compton telescope in which their high energy resolution is crucial for high angular resolution and background rejection capability. We have assembled a prototype module using a double-sided silicon strip detector and CdTe pixel detectors. 
In this paper, we present expected polarization performance of a proposed mission (NeXT/SGD).
We also report results from polarization measurements using polarized synchrotron light and validation of EGS4 MC simulation. 
\end{quote}

\vfill

\begin{center} 
{\it Contributed to} 
{\it UV-Gamma Ray Space Telescope Systems, Astronomical Telescopes and Instrumentation }\\
{\it Glasgow, UK}\\
{\it June 21--June 24, 2004} \\



\end{center}

\newpage



%
\pagestyle{plain}

\section{INTRODUCTION}
\label{sect:intro}  

The hard X-ray and gamma-ray bands have long been recognized as important windows for exploring the energetic universe. It is in these energy bands that non-thermal emission, primarily due to accelerated high energy particles, becomes dominant. However, by comparison with the soft X-ray band, where the spectacular data from the XMM-Newton and Chandra satellites are revolutionizing our understanding of the high-energy Universe, the sensitivities of hard X-ray missions flown so far, or currently under construction, have not dramatically improved over the last decade. Clearly, the scope of discovery expected with much improved sensitivity for both point and extended sources is enormous.

The energy band between 0.1~MeV and 100~MeV is poorly explored 
due to difficulties associated with the detection of such photons.
The Compton telescope COMPTEL~\cite{COMPTEL93} on board CGRO (Compton Gamma-Ray 
Observatory) demonstrated that a gamma-ray instrument based on the Compton 
scattering is useful for the detection of the gamma-ray in this energy band.
COMPTEL provided us rich information on a variety of gamma-ray emitting 
objects either in continuum and line emission. 
The continuum sources include spin-down pulsars, stellar black-hole candidates,
supernovae remnants, interstellar clouds, active galactic nuclei (AGN), 
gamma-ray bursts (GRB) and solar flares. 
Detection has also been made of the nuclear gamma-ray lines from 
${}^{26}$Al (1.809~MeV), ${}^{44}$Ti (1.157~MeV), and 
${}^{56}$Co (0.847 and 1.238~MeV). 

Although COMPTEL performed very well as the first Compton telescope 
in space for MeV gamma-ray astrophysics, it suffered severely 
from large background, poor angular resolution, and complicated image 
decoding.~\cite{Knodlseder96}
In 1987, T. Kamae {\it et al}. proposed a new Compton telescope based 
on a stack of silicon strip detectors (SSD)~\cite{Kamae87,Kamae88}.
This technology presents very attractive possibilities to overcome 
the weaknesses of COMPTEL as described later in this document.
This idea of using silicon strip detectors stimulated new proposals 
for the next generation Compton telescope~\cite{Takahashi02-NeXT,Takahashi03-SGD,MEGA,Milne}.

Recently, a new semiconductor detector based on Cadmium Telluride (CdTe) 
emerged as a promising detector technology for detection of MeV 
gamma-rays~\cite{Takahashi01,Nakazawa03}.
Taking advantage of significant development in CdTe technology, we are 
developing a new generation of Compton telescopes, the SGD (Soft Gamma-ray 
Detector)~\cite{Takahashi02-NeXT,Takahashi03-SGD,Takahashi04-SGD} onboard the NeXT (New X-ray Telescope) 
mission proposed at ISAS (Institute of Space and Astronautical Science) as a successor of the Astro-E2, and the SMCT (Semiconductor Multiple-Compton Telescope).

NeXT is optimized to study high-energy non-thermal processes: it will be a successor to the Astro-E2 mission, with much higher sensitivity in the energy range from 0.5 keV to 1 MeV.
NeXT will have a dramatic impact on the sensitivity for source detection at hard X-ray and soft gamma-ray energies.
The broad bandpass will allow us to determine the range of energies of the radiating non-thermal particles, but the measurement beyond 511 keV provides the next crucial step towards the determination of the source structure: this is because the rate of $e^+/e^-$ pair production and their annihilation and thus the flux of the 511 keV line - depends critically on the compactness of the source. This is where the currently envisioned Soft Gamma-ray Detector (SGD) excels, with dramatically lower background as compared to previous instruments.

The NeXT/SGD is a hybrid semiconductor gamma-ray detector which consists of 
silicon and CdTe detectors to measure photons in a wide energy band (0.05--1 MeV);
the silicon layers are required to improve the performance at a lower energy band ($<$0.3 MeV).
The silicon layers also improve the angular resolution because of smaller effect from the finite momentum of the Compton-scattering electrons (Doppler broadening) than CdTe. 
The basic SGD detector module's design is based on a narrow FOV (field-of-view) instrument which utilizes a Compton kinematics telescope to enhance its background rejection capabilities to achieve very low background ($10^{-7}$ \flux). 
The Compton kinematics rely on the excellent energy resolution afforded by the detector material in the SGD module. 
This resolution (better than 1.5 keV @ 60 keV) allows the instrument to achieve both high angular resolution and good background rejection capability.
As a natural consequence of the Compton approach used to decrease background, the SGD module is quite sensitive to X/$\gamma$-ray polarization, thereby opening up a new window to study particle acceleration mechanism in astronomical objects.

The SMCT will have a wider energy band (0.1--20 MeV), a large effective area ($\sim$1000~cm${}^2$) and a wide field of view ($\sim$60\degree).
Production and operating experiences with the SGD will be useful to improve the SMCT design.
In this paper, the polarization performance of NeXT/SGD is discussed mainly since more progress has been made on the SGD development rather than the SMCT.
\section{INSTRUMENT DESCRIPTION}

\begin{figure}[bth]
   \begin{center}
   \includegraphics[height=6cm]{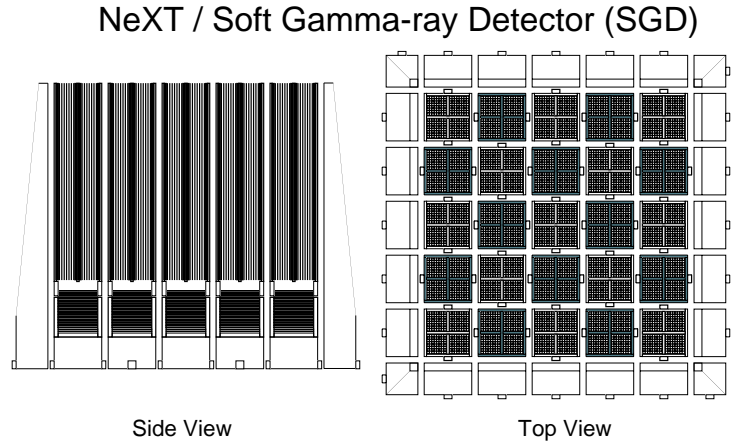}
   \end{center}
\vspace*{-0.4cm}
   \caption[Conceptual drawing of SGD.] 
   { \label{fig:SGD-drawing} Conceptual drawing of SGD.}
   \end{figure}

\begin{figure}[bth]
   \begin{center}
   \begin{tabular}{ll} 
   (a) & (b) \\
   \includegraphics[height=6cm]{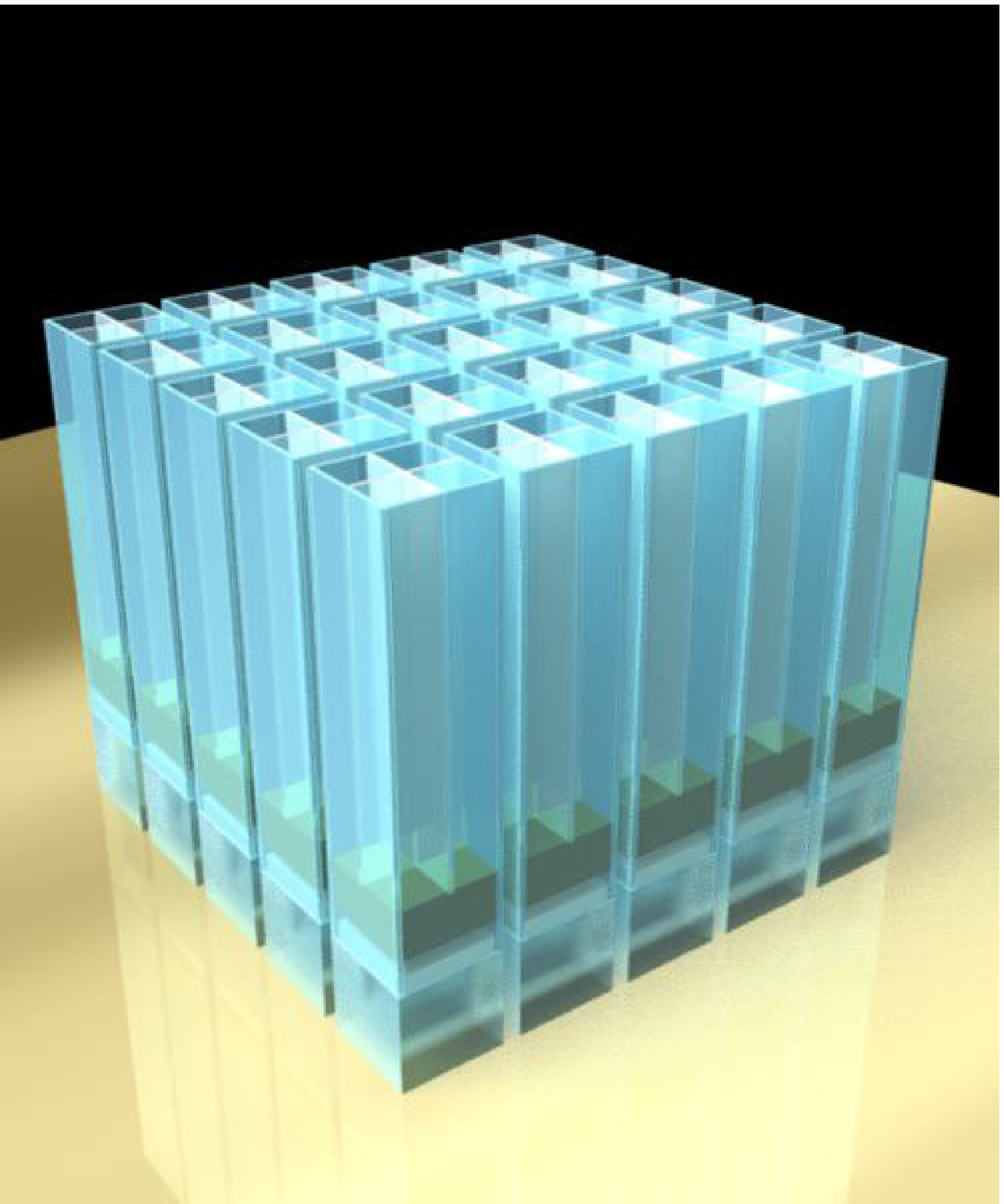} \hspace*{2cm} &
   \includegraphics[height=6cm]{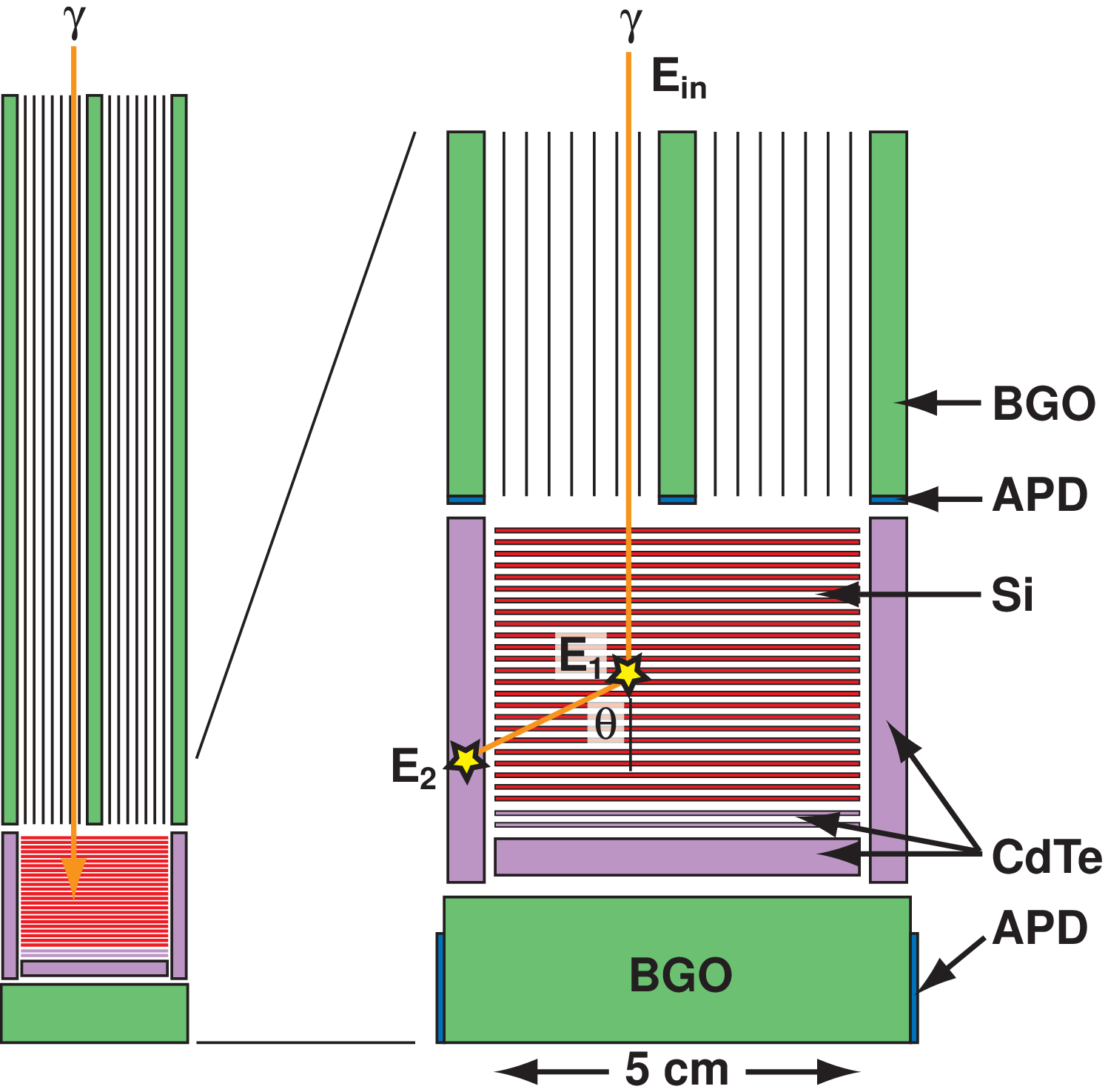}
   \end{tabular}
   \end{center}
\vspace*{-0.4cm}
   \caption[(a) 3D view of SGD detector units. (b) Conceptual drawing of a SGD detector unit.] 
   { \label{fig:SGD-unit} (a) 3D view of SGD detector units. (b) Conceptual drawing of a SGD detector unit.}
   \end{figure}

The NeXT/SGD is a Compton telescope with narrow FOV, which utilizes Compton kinematics to enhance its background rejection capabilities.
The hybrid design of this module, illustrated in Figure~\ref{fig:SGD-unit}, incorporates both silicon strips (to enhance response below $\sim $300 keV) and pixelated CdTe detectors.
The Compton telescope consists of 24 layers of DSSDs (double-sided silicon strip detectors) and 2 layers of thin CdTe (Cadmium Telluride) pixelated detectors surrounded by 5 mm thick CdTe pixelated detectors. 
A copper collimator restricts the field of view of the telescope to 0.5\degree\ for low energy photons ($<$100~keV), which is essential to minimize the CXB (cosmic X-ray backgrounds). The telescope is surrounded by BGO (Bi${}_4$Ge${}_3$O${}_{12}$) shield units. 
Scintillation light from a BGO crystal is detected by avalanche photo-diodes allowing a compact design. 
The combination of a Compton telescope, BGO active collimator, a fine passive collimator constructed of copper and a BGO bottom shield forms a detector module that can detect photons in a wide energy band (0.05--1 MeV) at a background level of $10^{-7}$ \flux).
These modules are then arrayed to provide the required area. Figure~\ref{fig:SGD-drawing} shows a conceptual drawing of the design goal of the SGD instrument, which consists of a $5\times 5$ array of identical detector modules.

We require each SGD event to interact twice in the stacked detector, once by Compton scattering in the Si part, and then by photo-absorption in the CdTe part. Once the locations and energies of the two interactions are measured, the Compton kinematics allows us to calculate the angle between the telescope axis and the incident direction of the event. The high energy resolution of the Si and CdTe devices help reduce the width of these ``Compton rings''. 
We can determine the location of point sources as intersections of multiple rings. The angular resolution is limited to $\sim$4\degree\ at 100 keV due to the Doppler broadening, which is comparable to the FOV of the BGO collimators. Although the order of the events can be uncertain, for the lower energies we can use the relation that the energy deposition by Compton scattering is always smaller than that of the photo absorption for energies below $E_{\gamma} = 256$ keV ($E_{\gamma} = m_e/2$). This relation holds above this energy, if the scattering angle $\theta$ is smaller than $\cos^{-1}(1 - m_e / 2E_{\gamma}$). The major advantage of employing the Compton kinematics, however, is to reduce the background. By having a narrow FOV, and by requiring the Compton ring of a valid aperture gamma-ray event to intersect with the FOV, we can reject most of the background events. This dramatically reduces the background from radio-activation of the detector materials, which is a dominant background term in the case of the Astro-E2 HXD (Hard X-ray Detector).\cite{HXD,HXD-BG}
Furthermore, we can eliminate Compton rings produced by bright sources located outside the FOV, which could produce significant background in some circumstances.

\section{Basic Polarization Performance}

The EGS4 Monte Carlo simulation package\cite{EGS4} with low energy extension\cite{EGS-KEK} is used to study the SGD performance.
We performed several measurements and verified its predictions\cite{Tajima03,Mitani03} as described in section~\ref{sect:exp}.
In this section, EGS4 simulation studies of basic SGD performance for the polarization are described.

\subsection{Expected Sensitivity}
\begin{figure}[bth]
   \begin{center}
   \begin{tabular}{ll} 
   (a)  & (b)  \\
   \includegraphics[height=5cm]{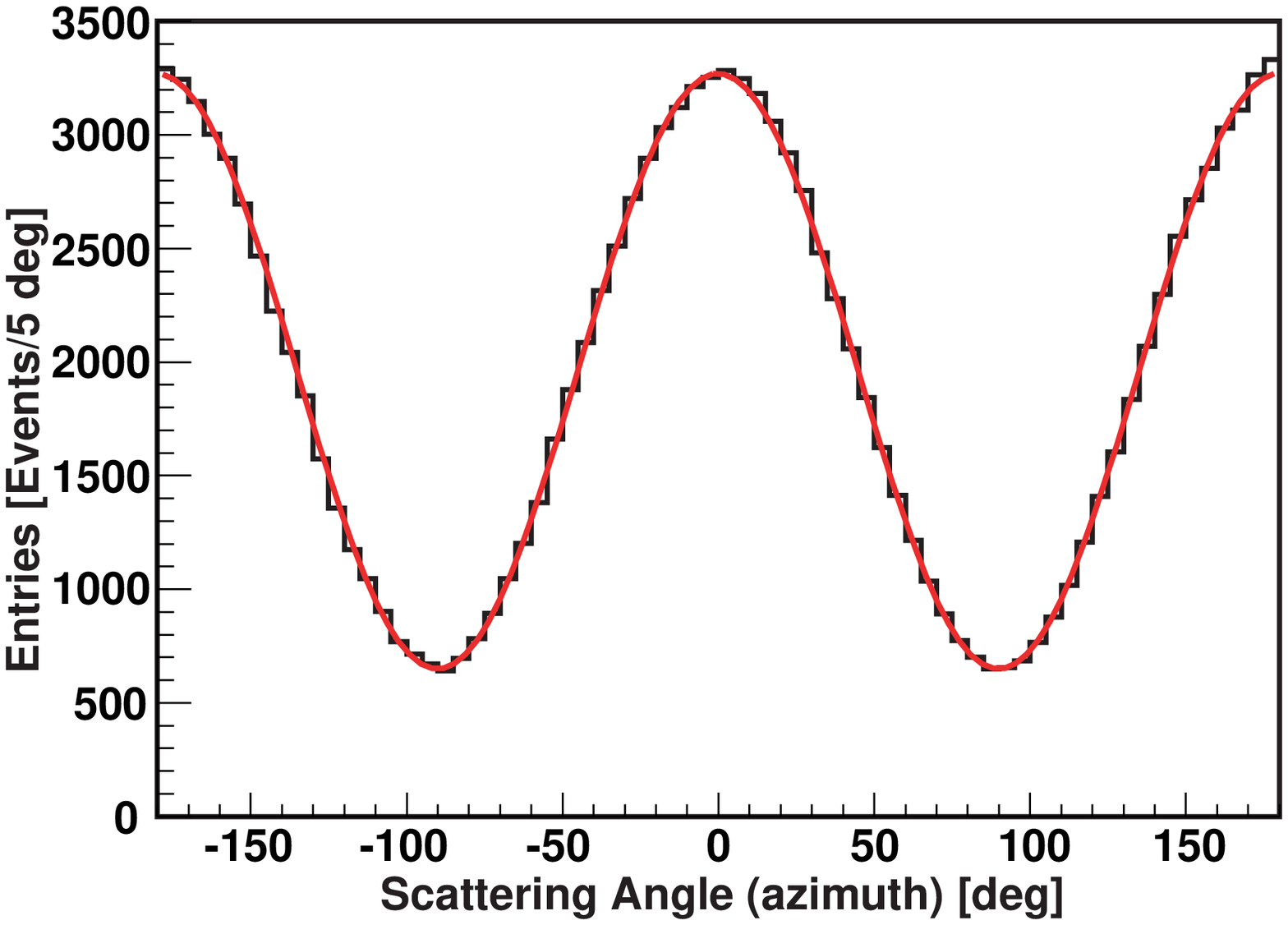} \hspace*{1.2cm} &
   \includegraphics[height=5cm]{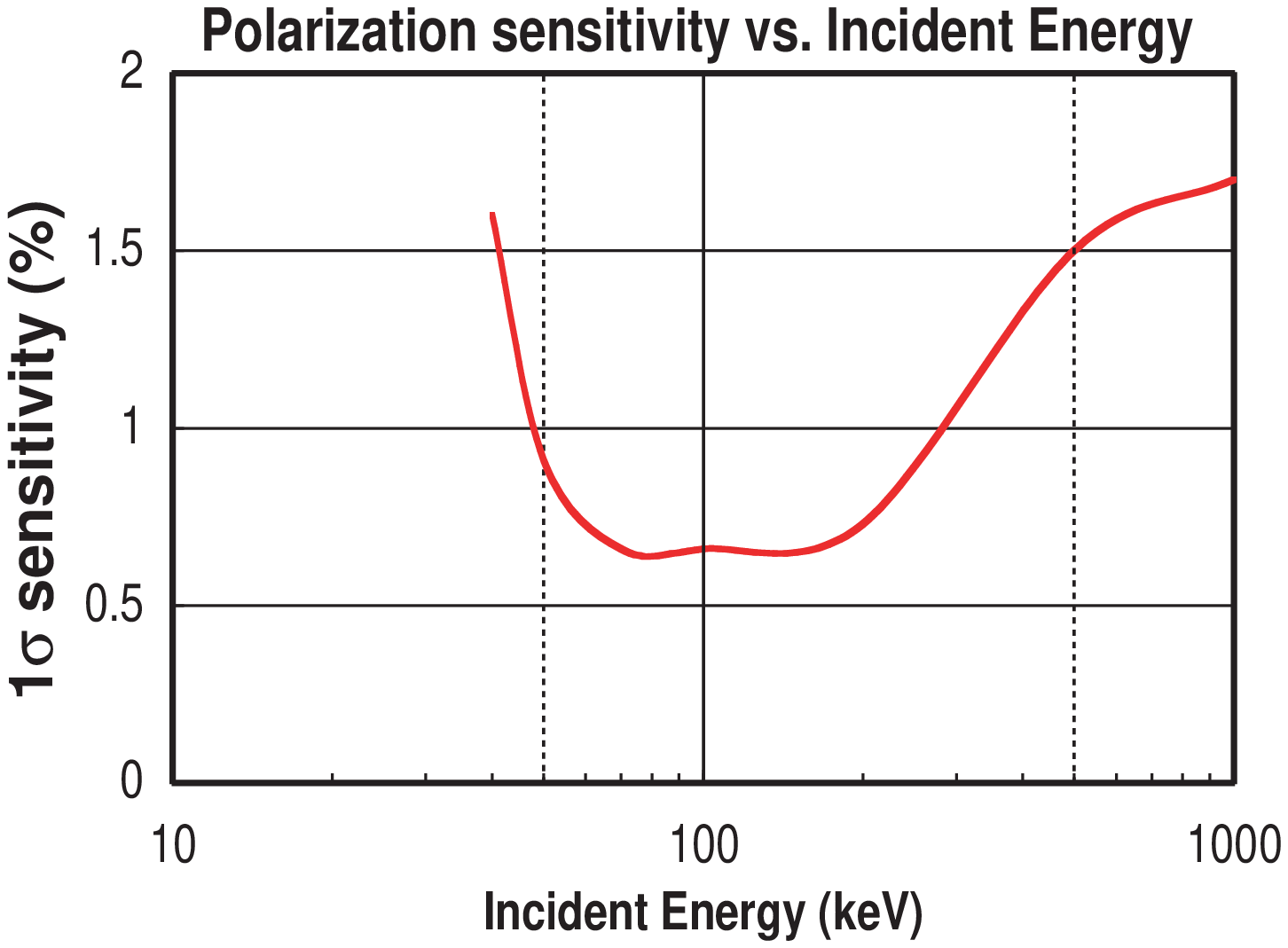}
   \end{tabular}
   \end{center}
\vspace*{-0.4cm}
   \caption[(a) Azimuth angle ($\phi$) distribution of the Compton scattering reconstructed for a 100~ks observation of 250 mCrab source with 100\% polarization.
   (b) 1$\sigma$ sensitivity for the linear polarization as a function of the incident energy for 0.5 million incident photons.]
   { \label{fig:pol-sensitivity}(a) Azimuth angle ($\phi$) distribution of the Compton scattering reconstructed for a 100~ks observation of 250 mCrab source with 100\% polarization.
   (b) 1$\sigma$ sensitivity for the linear polarization as a function of the incident energy for 0.5 million incident photons.}
   \end{figure}

We studied the sensitivity for the polarization with the Crab spectrum in the energy range 40--1000~keV using the EGS4 MC simulation.
Figure~\ref{fig:pol-sensitivity} (a) shows the azimuth angle ($\phi$) distribution of the Compton scattering reconstructed for a 100~ks observation of 150 mCrab source, $flux(E) = 1.3\cdot E^{-2.1}$~\flux, with 100\% polarization.\footnote{Polarization refers to linear polarization throughout this paper.}
The distribution is fit to a formula, $AVG\cdot (1+Q\cos2(\phi-\chi))$, where $AVG$ is the average flux per bin, $Q$ is the modulation factor which should be proportional to the polarization, and $\chi$ is the polarization phase and perpendicular to the polarization vector.
The fit yields the modulation factor ($Q$) of $66.89 \pm 0.30$\%, which corresponds to 1$\sigma$ polarization sensitivity of 0.45\%. 
It should be noted that good timing resolution of the SGD (better than 1 $\mu$s) enable us to study the pulsar polarization as a function of the pulse phase.
Figure~\ref{fig:pol-sensitivity} (b) shows the 1$\sigma$ sensitivity for the linear polarization as a function of the incident energy for 0.5 million incident photons.
It indicates that the SGD is most sensitive to the polarization between 60 and 200~keV.

\subsection{Systematic Bias}

\begin{figure}[bth]
   \begin{center}
   \begin{tabular}{ll} 
   (a)  & (b)  \\
   \includegraphics[height=5cm]{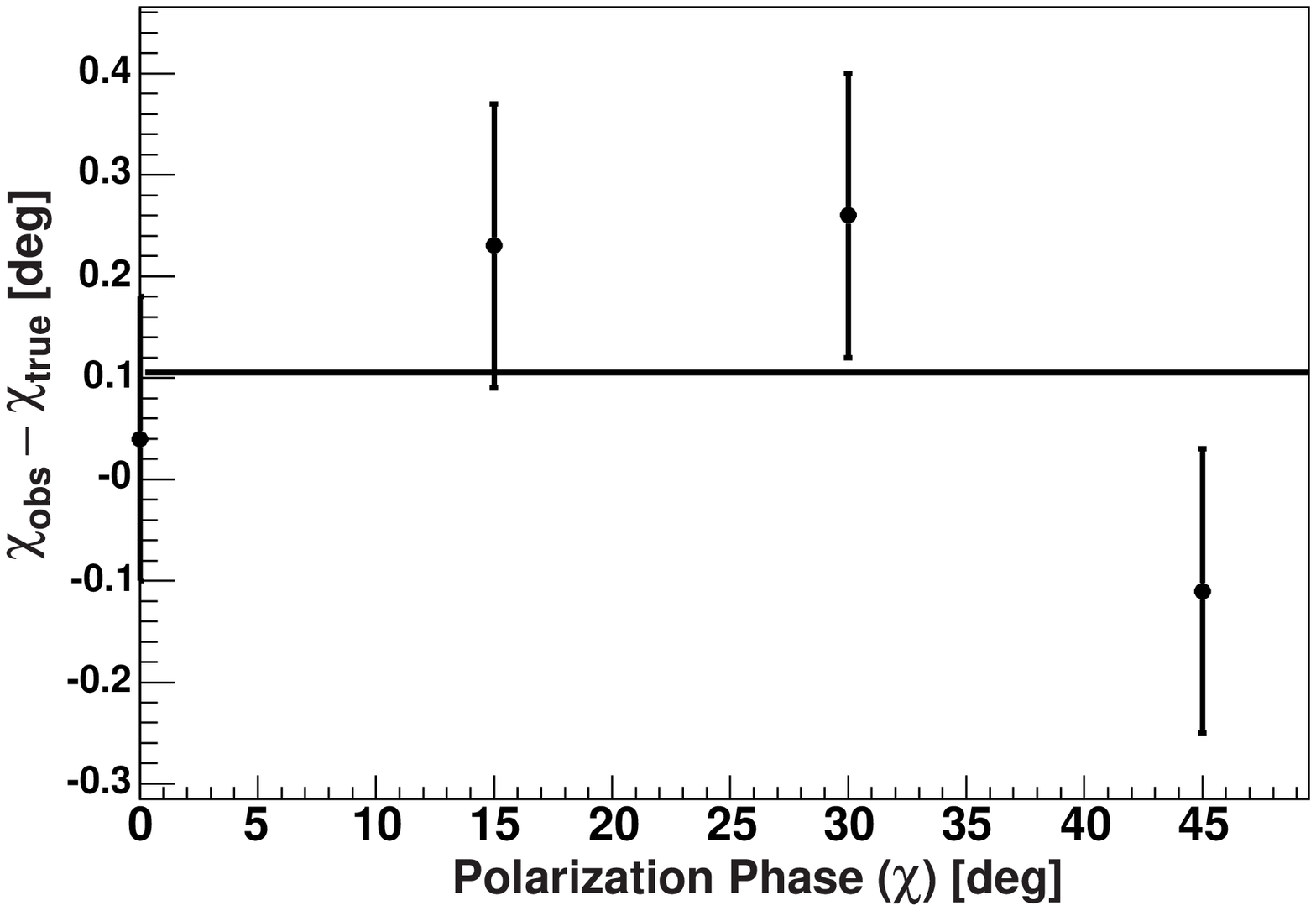} \hspace*{1.2cm} &
   \includegraphics[height=5cm]{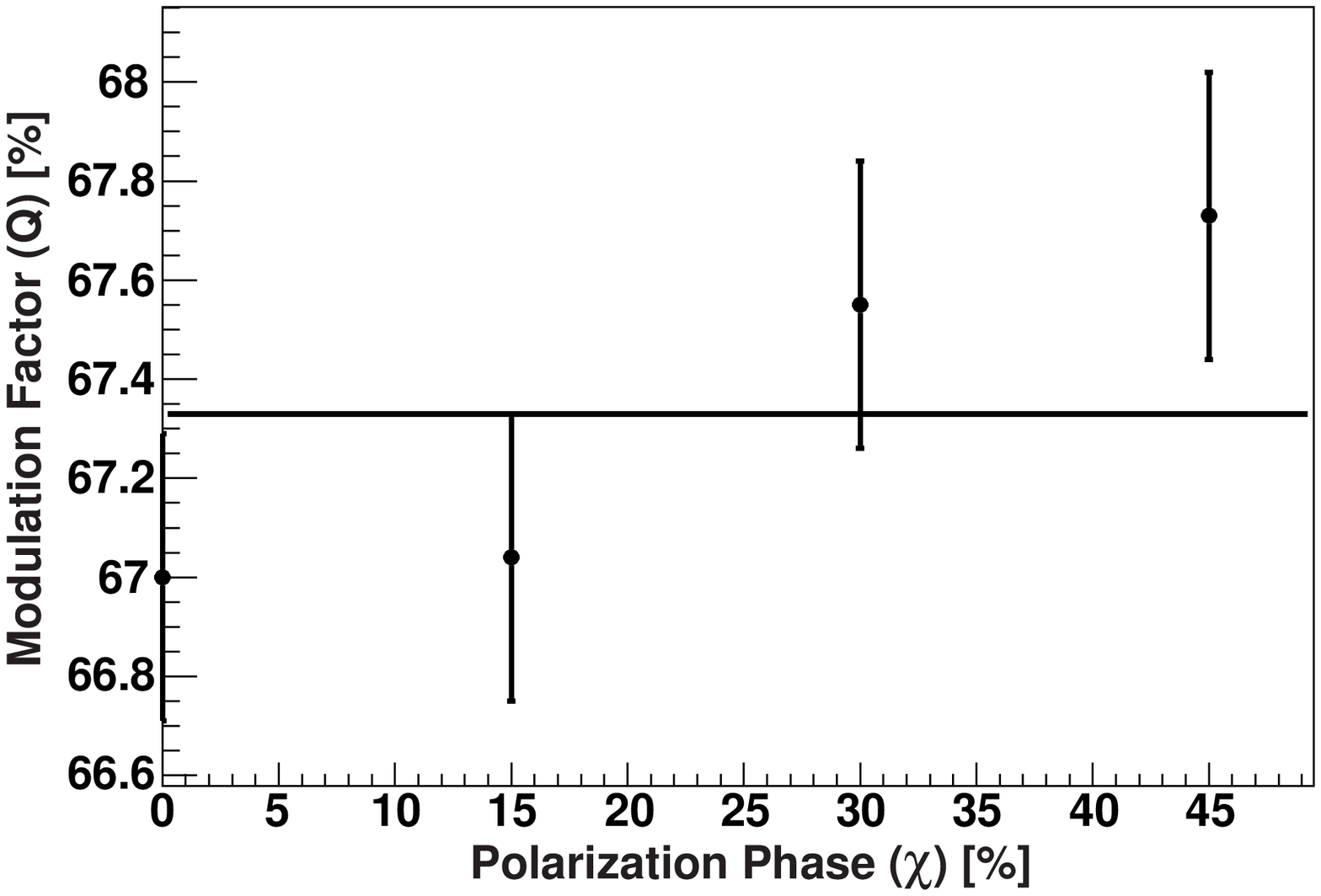}
   \end{tabular}
   \end{center}
\vspace*{-0.4cm}
   \caption[(a) The phase bias (observed phase minus true phase) and (b) the modulation factor ($Q$) as a function of the polarization phase with 100\% polarization.]
   { \label{fig:sys-phase}(a) The phase bias (observed phase minus true phase) and (b) the modulation factor ($Q$) as a function of the polarization phase with 100\% polarization.}
   \end{figure}

\begin{figure}[bth]
   \begin{center}
   \begin{tabular}{ll} 
   (a)  & (b)  \\
   \includegraphics[height=5cm]{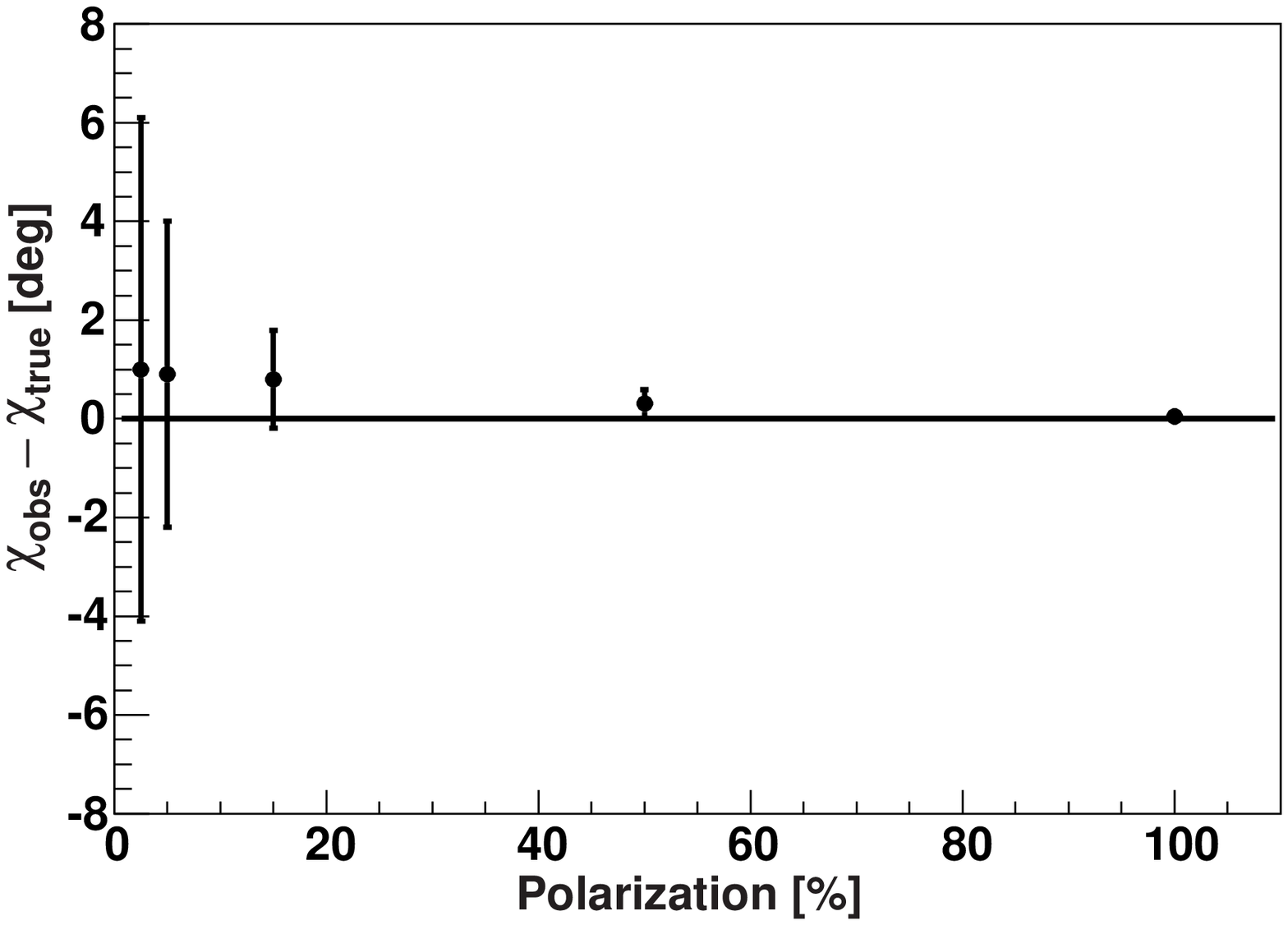} \hspace*{1.2cm} &
   \includegraphics[height=5cm]{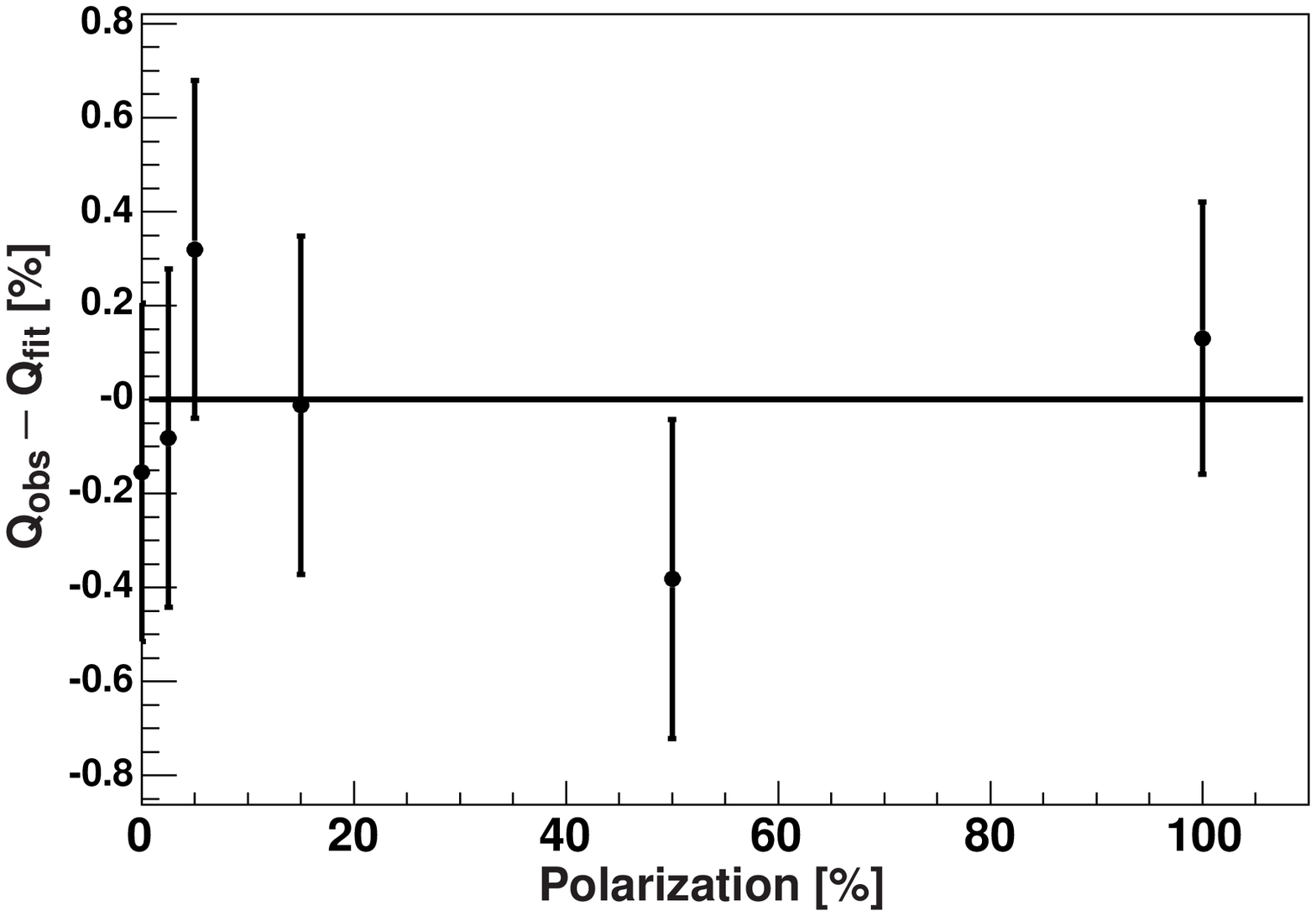}
   \end{tabular}
   \end{center}
\vspace*{-0.4cm}
   \caption[(a) The phase bias and (b) the $Q$ bias (observed $Q$ minus the linear fit of $Q$) as a function of the polarization of the incident photons.]
   { \label{fig:sys-pol}(a) The phase bias and (b) the $Q$ bias (observed $Q$ minus the linear fit of $Q$) as a function of the polarization of the incident photons.}
   \end{figure}

In this section, we search for possible systematic biases in polarization measurements primarily due to detector geometry.
Several samples are generated by the EGS4 MC simulation for different polarization degree and phase to evaluate the bias depending on these parameters.
Figures~\ref{fig:sys-phase} (a) and (b) show the phase bias, observed phase ($\chi_\mathrm{obs}$) minus true phase ($\chi_\mathrm{true}$),
and the modulation factor ($Q$) as a function of the polarization phase with 100\% polarization.
It appears the phase bias is negligible while the $Q$ bias could be in the
order of 1\%.
The $Q$ bias seems to be maximum when the polarization phase is 45\degree, {\it i.e.} diagonal to the instrument.
Figure~\ref{fig:sys-pol} (a) shows the phase bias as a function of the polarization of the incident photons.
Slight bias can be noticed at low polarization, however it is much smaller
than the statistical error.
It also shows that the error of the phase measurement is 5\degree\ at the 5$\sigma$ detection limit, {\it i.e.} 2.5\% polarization.
Figure~\ref{fig:sys-pol} (d) shows the difference between the observed $Q$ ($Q_\mathrm{obs}$) and that from the linear fit of $Q$ ($Q_\mathrm{fit}$) as a function of the polarization.
It indicates that the $Q$ linearity is better than 1\%.

\begin{figure}[bth]
   \begin{center}
   \begin{tabular}{ll} 
   (a)  & (b)  \\
   \includegraphics[height=5cm]{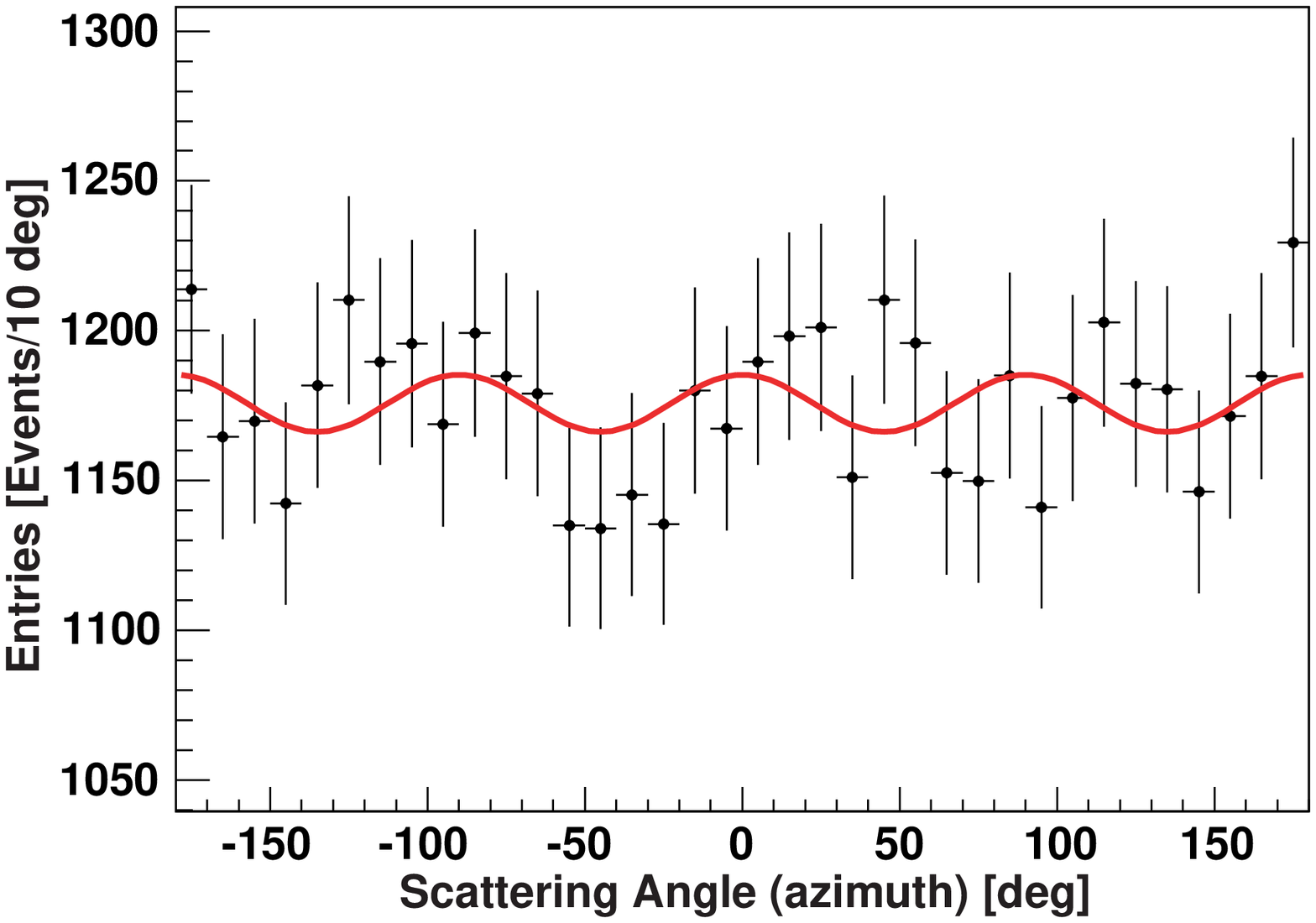} \hspace*{1.2cm} &
   \includegraphics[height=5cm]{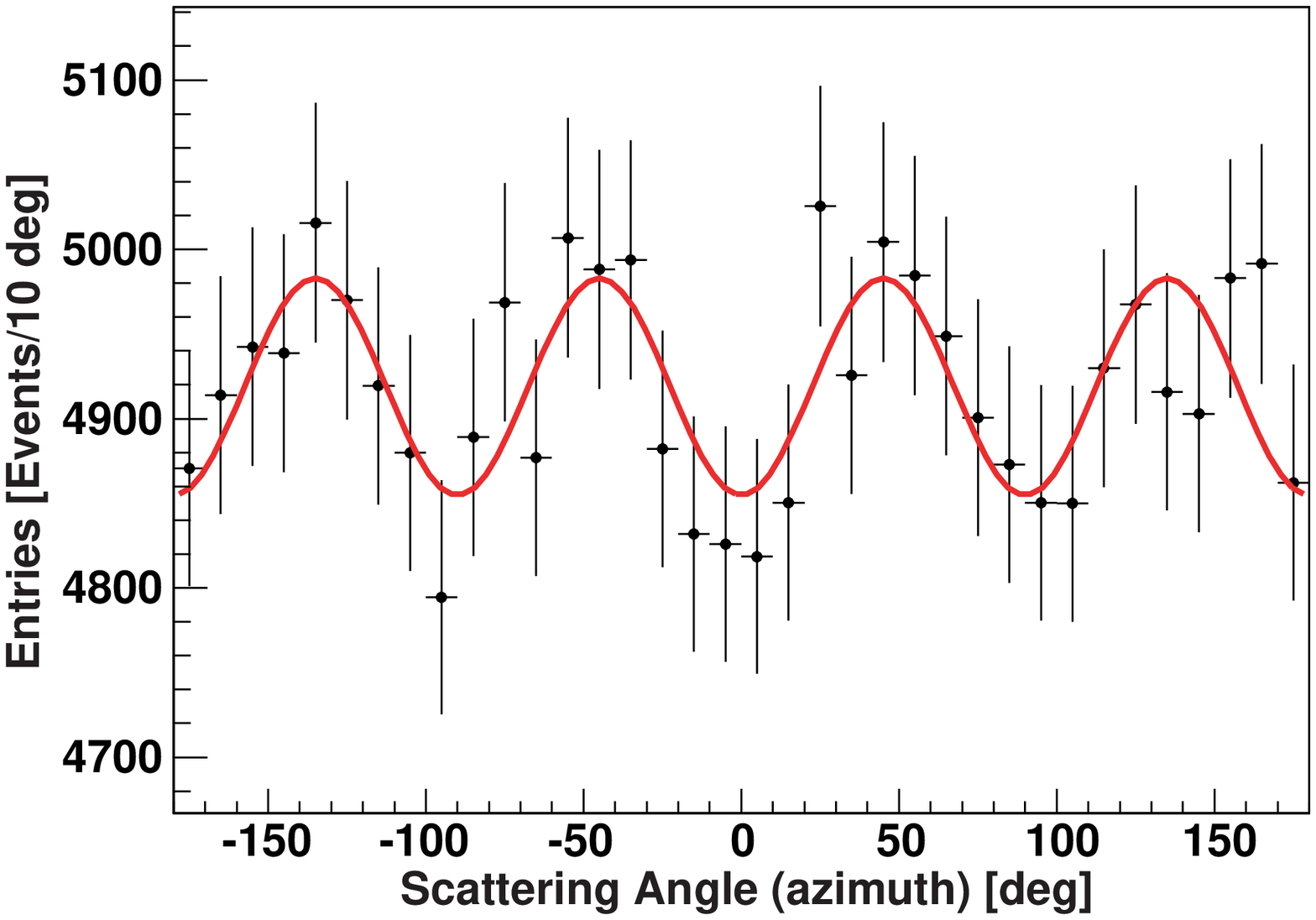} \\
   (c)  & (d)  \\
   \includegraphics[height=5cm]{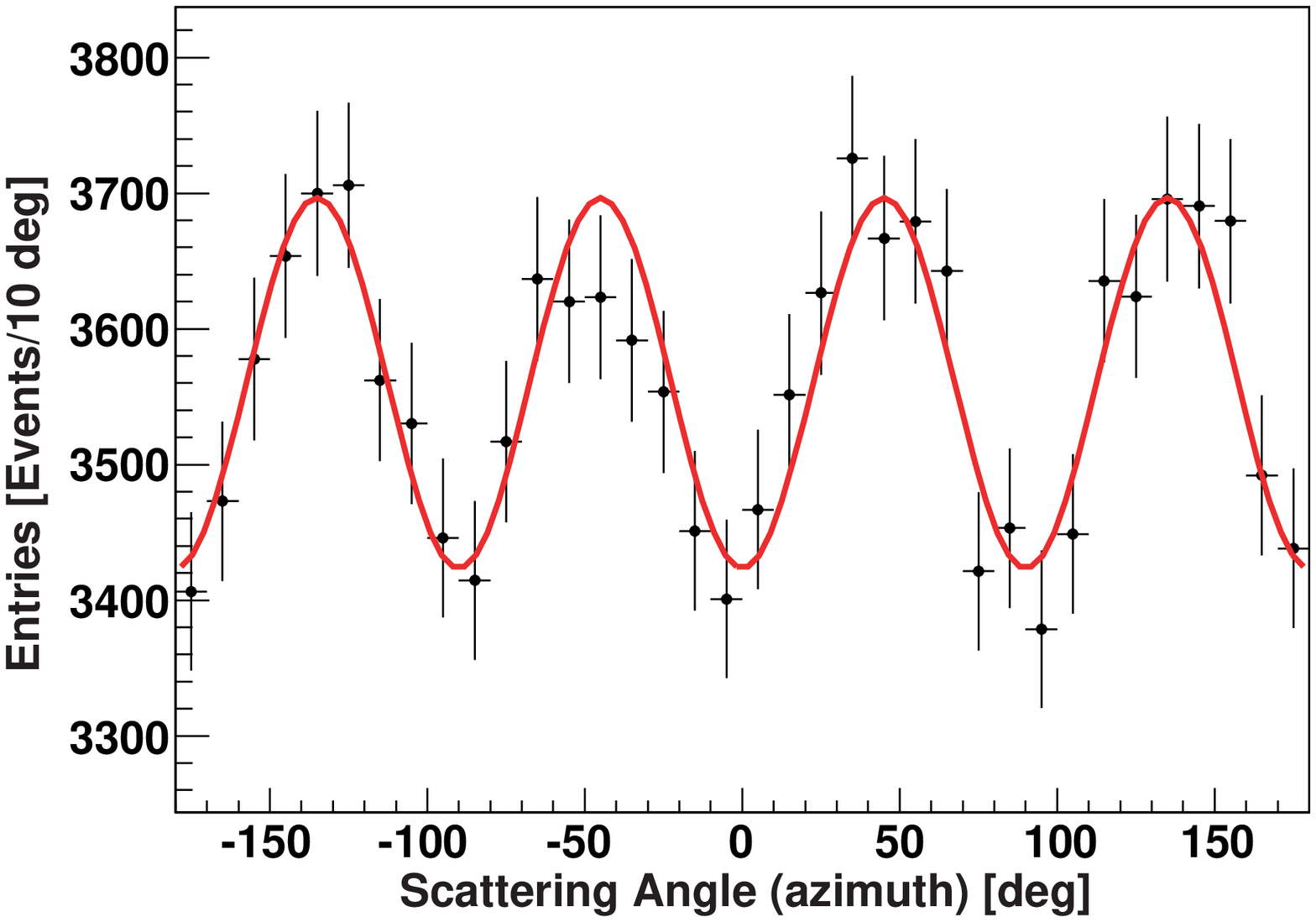} \hspace*{1.2cm} &
   \includegraphics[height=5cm]{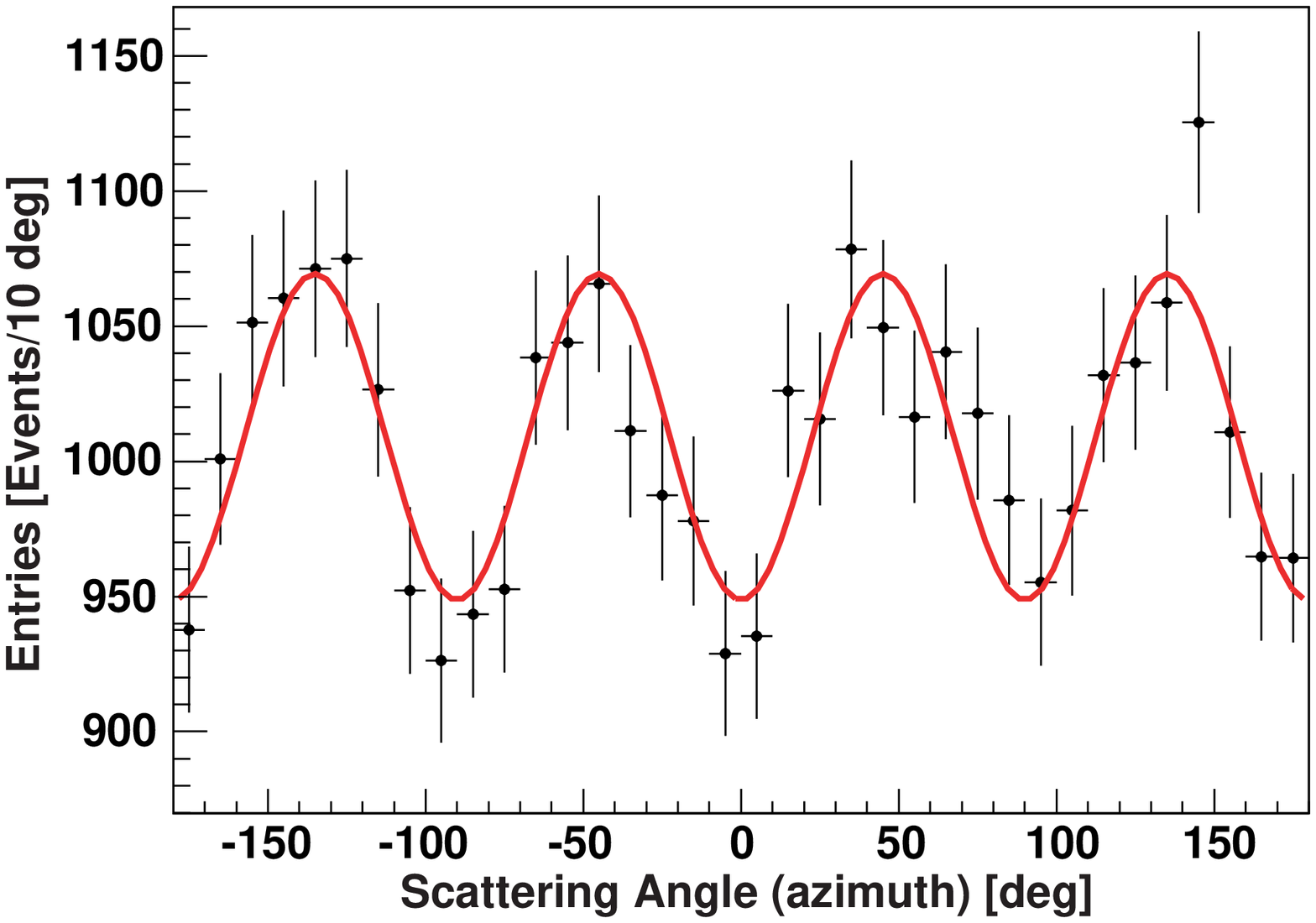}
   \end{tabular}
   \end{center}
\vspace*{-0.4cm}
   \caption[(a) Azimuth scattering angle distributions with no polarization for (a) $<$100 keV, (b) 100--300 keV, (c) 300--600 keV, (d) $>$600 keV simulated by the EGS4 MC simulation.]
   { \label{fig:sys-nopol}Azimuth scattering angle distributions with no polarization for (a) $<$100 keV, (b) 100--300 keV, (c) 300--600 keV, (d) $>$600 keV simulated by the EGS4 MC simulation.}
   \end{figure}

We have also simulated the response to unpolarized photons.
Figure~\ref{fig:sys-nopol} shows the azimuth angle distributions with no polarization for (a) $<$100 keV, (b) 100--300 keV, (c) 300--600 keV, (d) $>$600 keV.
We observe four bumps due to the the square arrangement of the CdTe wall in the base module.
These bumps correspond to the effective thickness of the CdTe wall for scattered photons.
In order to make it clear such effect, we consider the case in which the Compton scattering takes place at the center of the base module.
The absorption efficiency of the scattered photon in the CdTe wall, $\epsilon$, can be expressed as 
\begin{eqnarray}
\epsilon =& 1-e^{-\frac{t}{\lambda(E)|\sin\theta\cos\phi|}};\;\phi<-135,\;-45<\phi<45,\;\phi>135,\\
 & 1-e^{-\frac{t}{\lambda(E)|\sin\theta\cos(\phi-90)|}};\;-135<\phi<-45,\; 45<\phi<135,
\end{eqnarray}
where $t$ is the thickness of the CdTe wall, $\lambda$ is the photon absorption length of CdTe,
$E$ is the photon energy, and $\theta$ and $\phi$ are the polar and azimth scattering angles.
At low energy, $\epsilon\approx1$ becuase $t >> \lambda(E)$.
At high energy, this function peaks at $\phi=-135,\;-45,\;45,\;135$\degree.
The effect is larger at high energy since the detector become transparent.
The $\phi$ distribution in Figure~\ref{fig:sys-nopol} can be approximated as $AVG\cdot(1-Q_B\cos4\phi)$.
Fit to this formula yield (a) $Q_B=-0.005\pm0.005$, (b) $Q_B=0.015\pm0.005$, (c) $Q_B=0.035\pm0.005$ and (d) $Q_B=0.065\pm0.005$.
Note this effect does not produce fake polarization due to 4-fold asymmetry.
These effects are taken into account in the plots shown below.

These results suggest that SGD has no serious systematic bias on the polarization measurement.

\section{Expected Scientific Polarization Performance}
In this section, scientific performance of the SGD is discussed.
The EGS4 MC simulation is performed for selected astronomical objects where some degree of polarization is expected.
Since there is no definite prediction for the magnitude of the polarization of hard X-ray band in most objects, here we attempt to establish the 5$\sigma$ detection limit of the polarization for each object.

\subsection{Galactic Black Holes}

\begin{figure}[bth]
   \begin{center}
   \begin{tabular}{ll} 
   (a)  & (b)  \\
   \includegraphics[height=5cm]{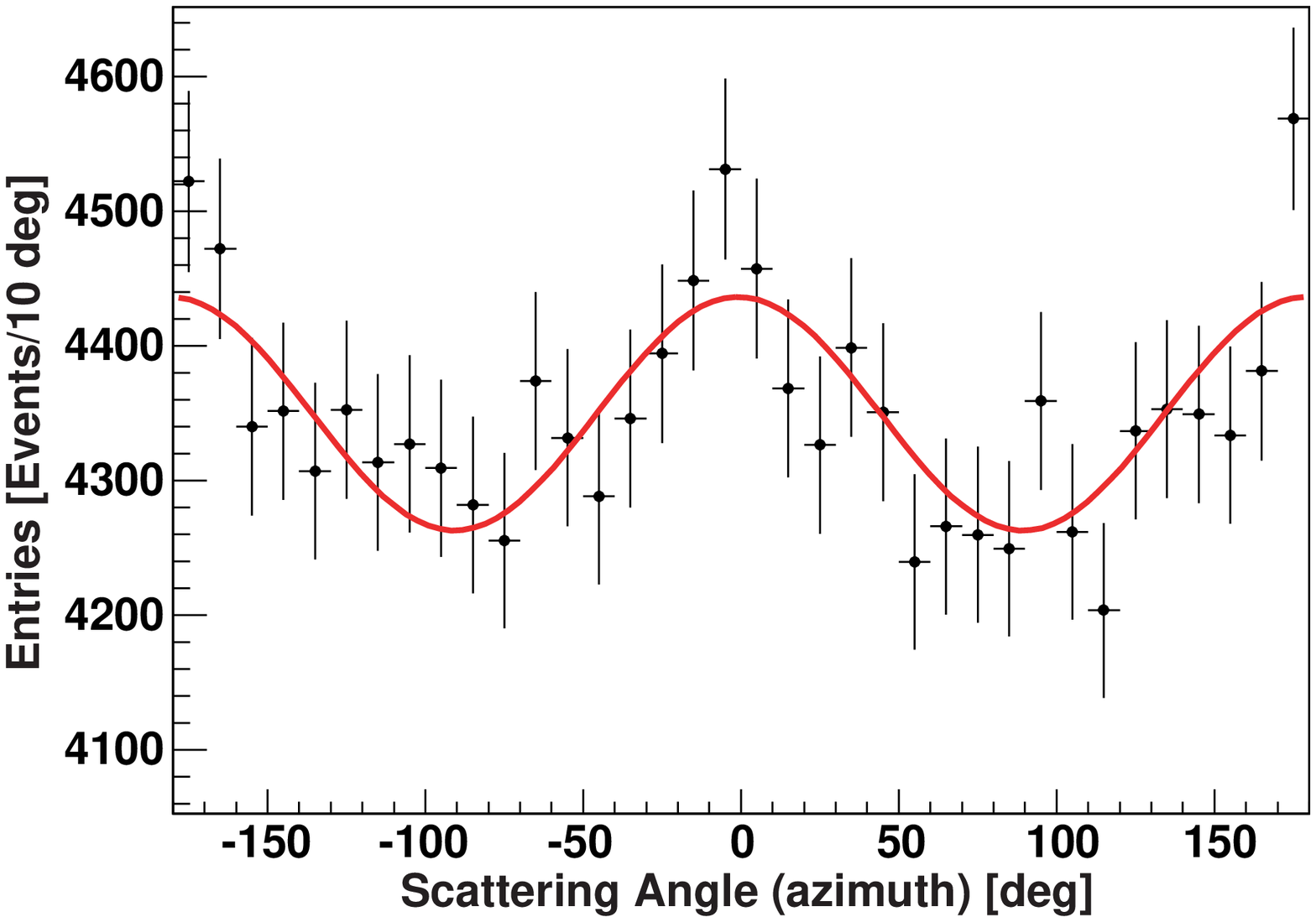} \hspace*{1.2cm} &
   \includegraphics[height=5cm]{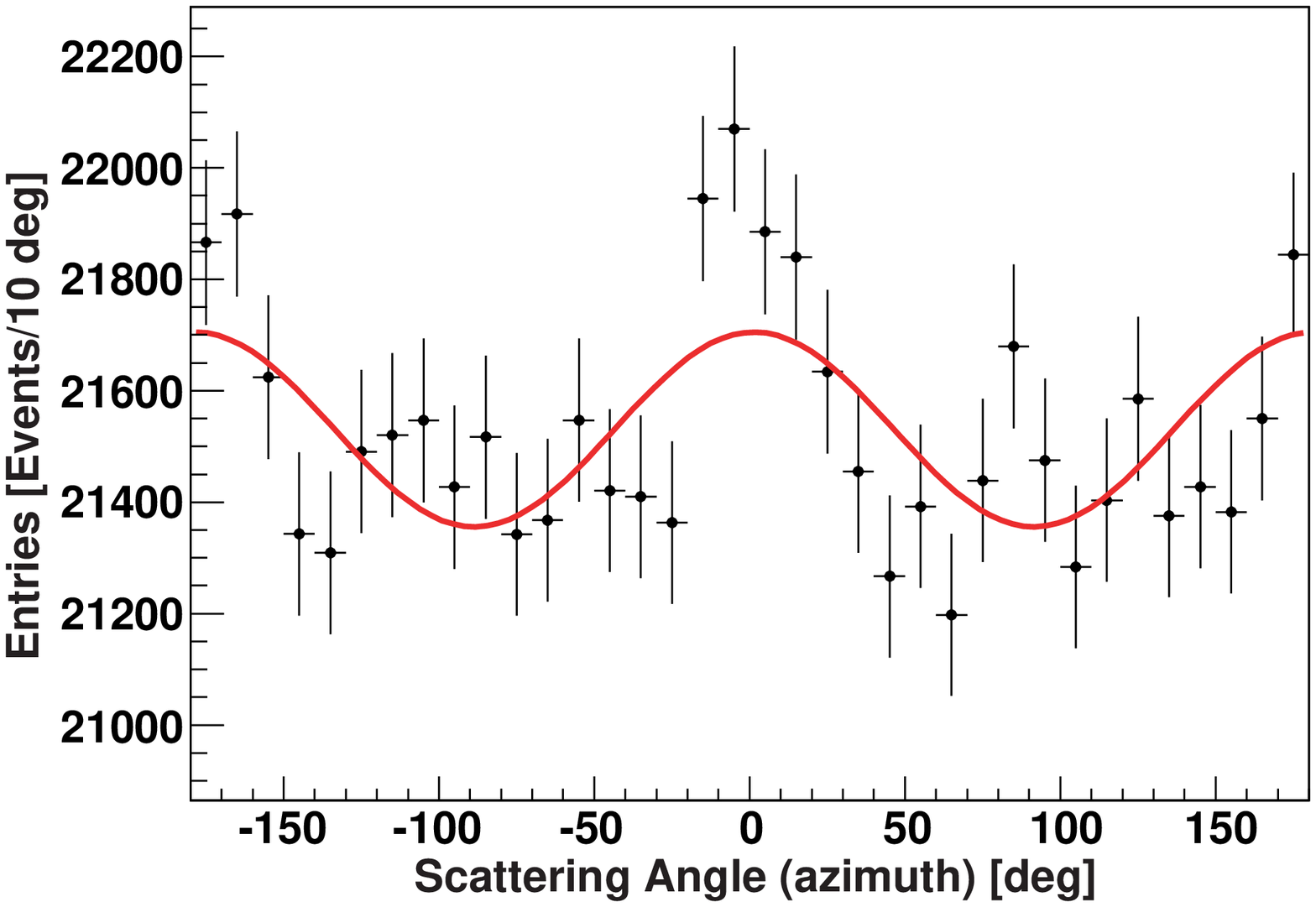}
   \end{tabular}
   \end{center}
\vspace*{-0.4cm}
   \caption[Azimuth Compton scattering angle distribution in the SGD at 5$\sigma$ detection limit with 100~ks observations of Cygnus X-1 in (a) the soft state and (b) hard state.]
   { \label{fig:CygX1-pol} Azimuth Compton scattering angle distribution in the SGD at 5$\sigma$ detection limit with 100~ks observations of Cygnus X-1 in (a) the soft state and (b) hard state.}
   \end{figure}

Polarization measurement of radiation from accreting Galactic black holes such as Cygnus X-1 is essential to understand the photon emission mechanism in the hard X-ray band, and to constrain the source geometry.  In the soft state, most models assume that hard X-rays originate from single Compton scattering by very hot, relativistic plasma present near the accretion disk.
Since the disk is inclined to the line of sight, large polarization is expected.  In the hard state, the hard X-ray emission is also due to Compton upscattering of soft photons, but the plasma has lower temperature, and to produce hard X-ray photons, multiple Compton upscatterings are necessary.
Since in multiple-order Compton scatterings the information about the polarization of the original photon is lost, the primary hard X-ray flux is unlikely to be polarized.
However, the Compton reflection of the primary radiation from the accretion disk might have some degree of polarization, but the total level of polarization most likely will be lower than for Cygnus X-1 in the soft state.  
In any case, the electrical vector of scattered photons that are detected by an observer seeing such an inclined disk is most likely predominantly parallel to the plane of the accretion disk.
On the other hand, some models involve a jet as a source of hard X-rays and predict small polarization.
It is possible to determine the origin of hard X-rays (accretion disk or jet) using the degree and phase of the polarization.
Figure~\ref{fig:CygX1-pol} shows the azimuth Compton scattering angle distribution in the SGD with 100~ks observations of Cygnus X-1 in (a) the soft state (2.5\% polarization) and (b) hard state (1.2\% polarization).
Fits yields $Q=0.0200\pm0.0036$ and $Q=0.0081\pm0.0016$, corresponding to the polarization of $2.99\pm0.54$\% for the soft state and $1.23\pm0.24$\% for the hard state. 
($Q=0.6700\pm0.0029$ and $Q=0.658\pm0.0018$ for 100\% polarization in the soft and hard states, respectively. Note that the modulation factor depends on the energy spectrum.)
These figures clearly demonstrate that the SGD can detect 2.5\% in the soft state and 1.2\% polarization in the hard state at better than 5$\sigma$ significance.
If the SGD does not detect the polarization or finds that the polarization vector is not parallel to the accretion disk in the soft state, the corona model requires serious modification and the jet model become viable.

\subsection{Active Galactic Nuclei}
Our current best picture of Active Galactic Nuclei has the primary source of radiative power coming from gravity.
This occurs by accretion of interstellar material onto a black hole with a mass upwards of $10^6$ Solar masses via a flattened structure known as the accretion disk. 
In some cases, the accretion process is also associated with a formation of a powerful jet, moving at relativistic speed away from the black hole.
If such a jet points at us, the Doppler boost makes the jet radiation dominant to the point that the isotropic (non-jet) radiation is barely detectable; such sources are known as blazars.
The nature of such jets and their relationship to the accretion structure is one of the greatest questions in astrophysics.
With this, the study of the structure of AGN in general involves the study of the isotropic component in jet-less sources, and separately, the study of the structure and composition of the jets in blazars.
Blazars emit strongly over all observable electromagnetic bands, with the most spectacular detections of gamma-rays with EGRET and ground-based atmospheric Cerenkov instruments: in fact, the observed flux in hard gamma-rays in many sources dominates over other bands.
They are the brightest hard gamma-ray sources in the sky, and thus the most extreme laboratories for testing the particle acceleration scenarios.
Overall spectra of blazars consist of two broad components: one in radio/IR/optical, presumably produced via synchrotron process.
This is inferred from the non-thermal shape of their spectra, and polarization detected in all bands where it can be measured.
The other component, in hard X-rays / gamma-rays, is presumably due to the inverse Compton process by the same particle population as the synchrotron emission, with the target photons that are either internal to the jet (synchrotron) or external, due to the emission line regions or diffuse IR flux in the host galaxy\cite{Sikora03}. 
For the few blazars where the synchrotron spectra extend to a few hundred keV (such as the well-known Mkn 501,\cite{Kataoka98} or recently-discovered TeV emitter 1959+65\cite{Krawczynski}), detection of high polarization is nearly assured by theory.
A lack of hard X-ray polarization will seriously undermine current models for X-ray emission from TeV blazars.
In the sources where the hard X-rays are due to Compton scattering, measurement of  polarization will reveal the origin of the gseedh photons for Comptonization.
Specifically, if the ``seed" photons are already polarized, which would be the case if they are of synchrotron origin, (as in Synchrotron-self Compton models), the Compton emission (which is optically thin) will also be polarized, but if the ``seeds" are unpolarized (external radiation models), we do not expect any polarization.
This provides a clear discrimination of popular models. 

   \begin{figure}[bth]
   \begin{center}
   \begin{tabular}{ll} 
   (a)  & (b)  \\
   \includegraphics[height=5cm]{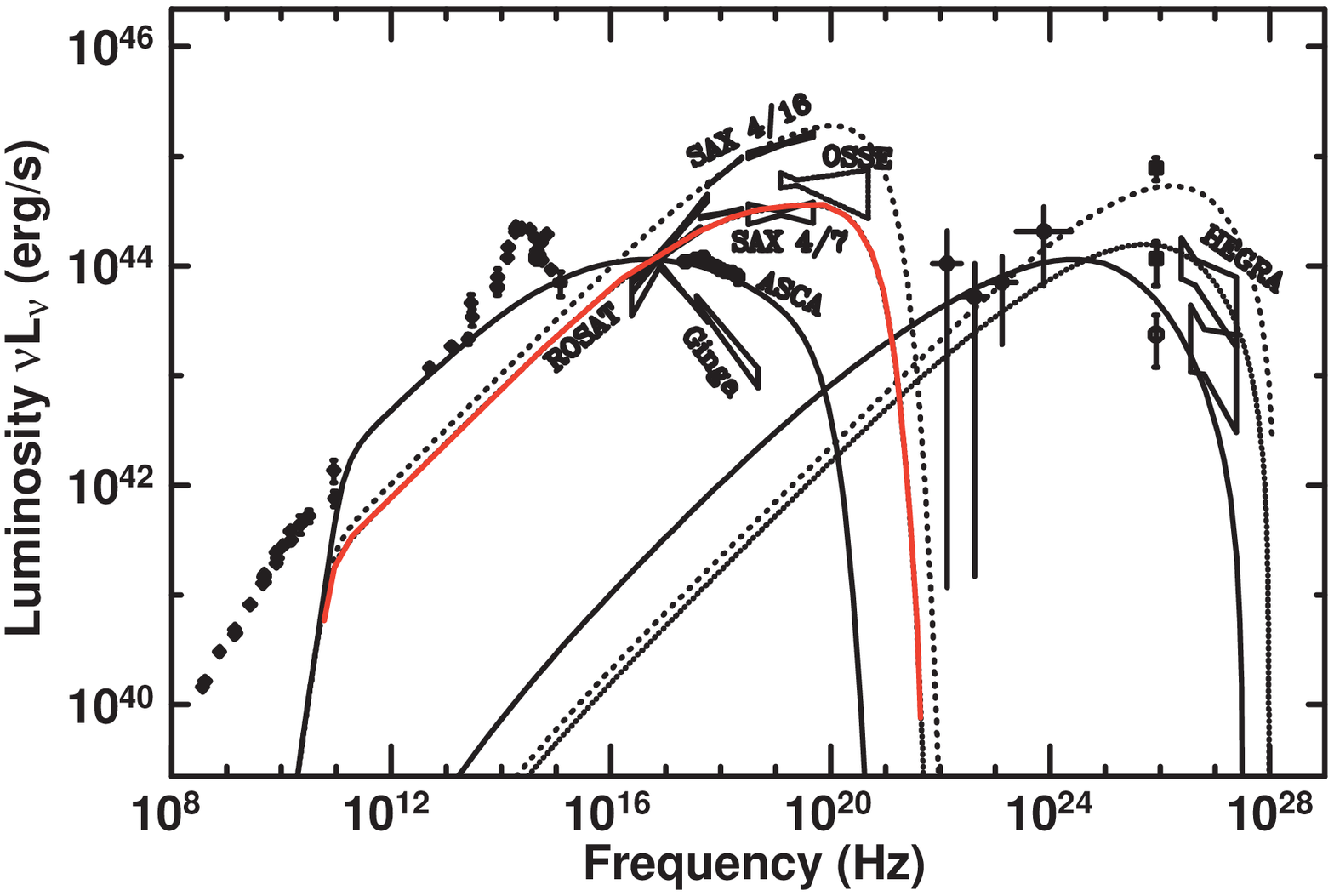} \hspace*{1.2cm} &
   \includegraphics[height=5cm]{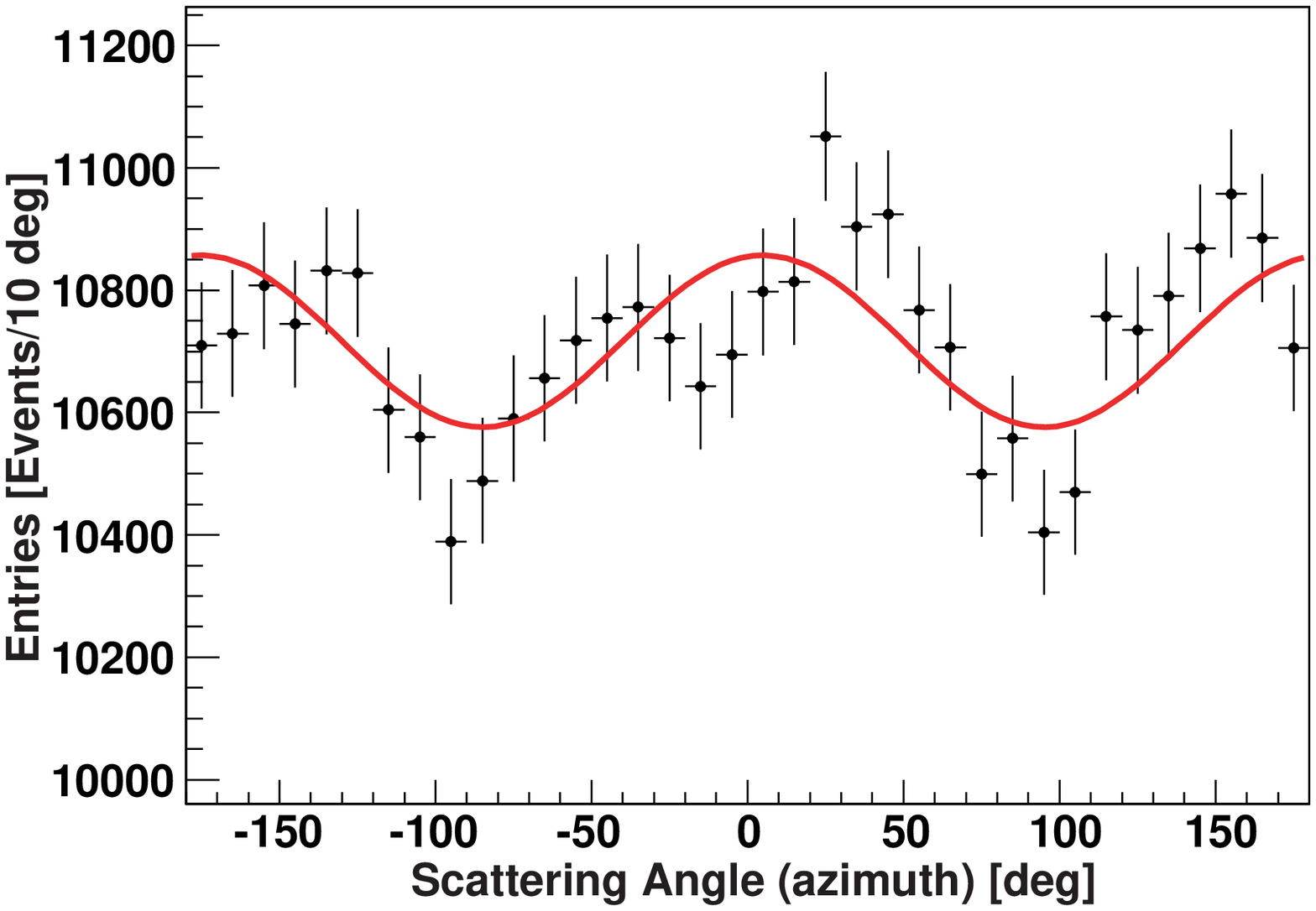}
   \end{tabular}
   \end{center}
\vspace*{-0.4cm}
   \caption[(a) Observed Mkn 501 Spectra. 
   (b) Simulated azimuth Compton scattering angle distribution in the SGD at 5$\sigma$ detection limit with 100~ks observations of Mkn 501.]
   {\label{fig:Mkn501-pol}(a) Observed Mkn 501 Spectra. 
   (b) Simulated azimuth Compton scattering angle distribution in the SGD at 5$\sigma$ detection limit with 100~ks observations of Mkn 501.}
   \end{figure} 
Figure~\ref{fig:Mkn501-pol} (a) shows the observed spectra of Mkn 501 (from Ref.~\citenum{Kataoka98}). 
The intermediate state is simulated here.
Figure~\ref{fig:Mkn501-pol} (b) shows the azimuth Compton scattering angle distribution in the SGD for 2.1\% linear polarization with 100~ks observations of Mkn 501.
We obtain $Q=0.0131\pm0.0023$, corresponding to the polarization degree of $2.26\pm0.39$\%. ($Q=0.5796\pm0.0024$ for 100\% polarization.)
It clearly demonstrates that the SGD can detect 2\% polarization at 5$\sigma$ significance.
Since high polarization ($>$10\%) is nearly assured in these objects by theory, non-detection of the polarization by the SGD would put many current models of blazars into question.

\subsection{Gamma-Ray Bursts}
Gamma-Ray bursts (GRBs), short and intense bursts of soft gamma-rays arriving from random directions in the sky, is considered to be the most luminous objects in the universe.
During the last decade, several space missions: BATSE (Burst and Transient Source Experiment) on Compton Gamma-Ray Observatory, BeppoSAX and now HETE II (High-Energy Transient Explorer), together with ground optical, infrared and radio observatories have revolutionized our understanding of GRBs.
BATSE has demonstrated that GRBs originate at cosmological distances in the most energetic explosions in the Universe. 
BeppoSAX discovered X-ray afterglow, enabling us to pinpoint the positions of some bursts, locate optical and radio afterglows, identify host galaxies and measure redshifts to some bursts.
It is a common understanding that some of GRBs are associated with core collapse Supernovae.

Recently, the Ramaty High Energy Spectroscopic Imager (RHESSI) reported that the prompt emission of the bright GRB 021206 was strongly polarized, and the claimed polarization degree is $80\pm20$\%\cite{CB03}.
Unfortunately, due to its close angular distance from the Sun (which actually facilitated the detection by RHESSI) prevented the detection of the optical afterglow.
In general, it is theoretically difficult to explain such strong polarization.
More sensitive measurements are essential to further understand the polarization properties of the GRBs.

\begin{figure}[bth]
   \begin{center}
   \begin{tabular}{ll} 
   (a)  & (b)  \\
   \includegraphics[height=5cm]{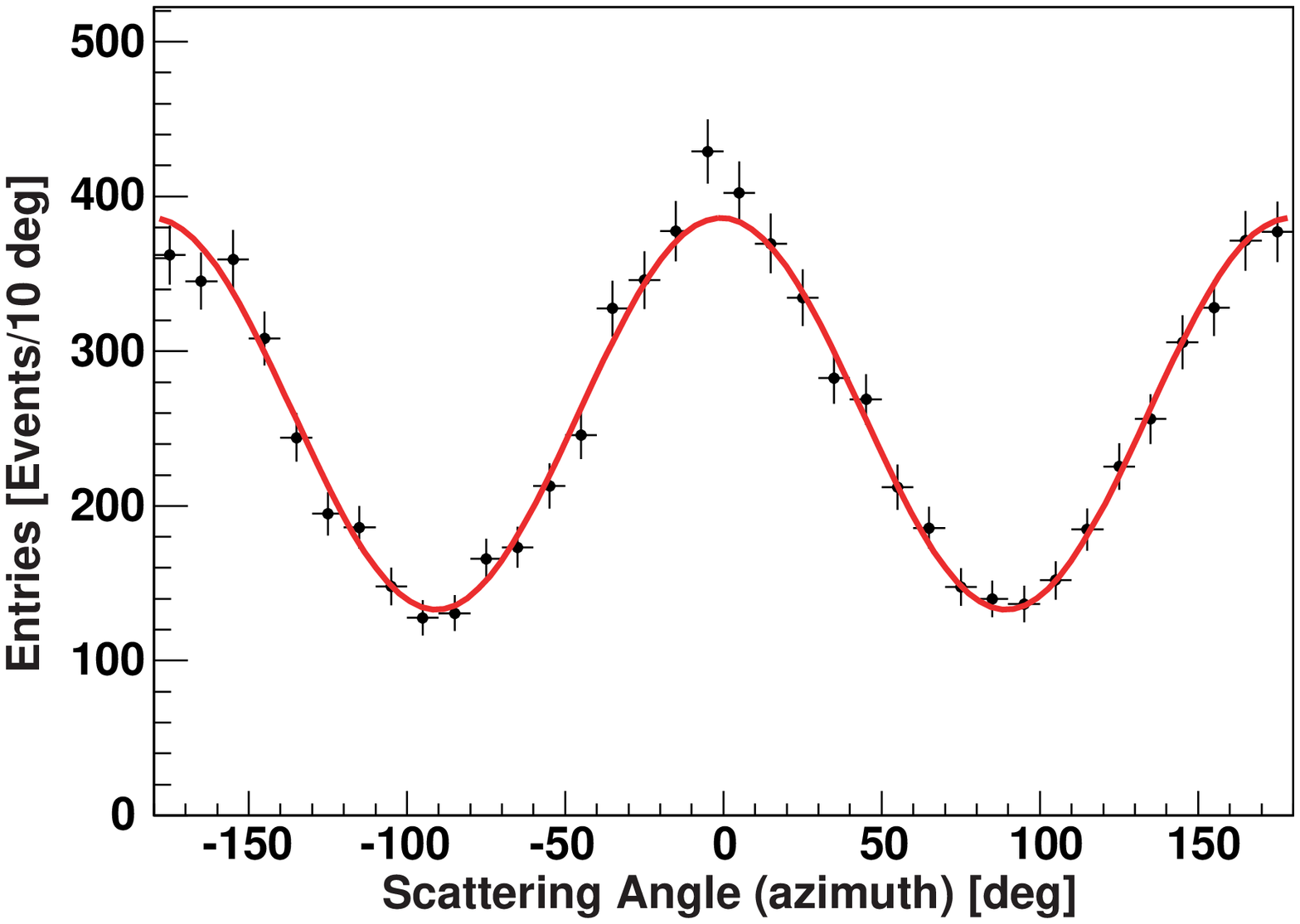} \hspace*{1.2cm} &
   \includegraphics[height=5cm]{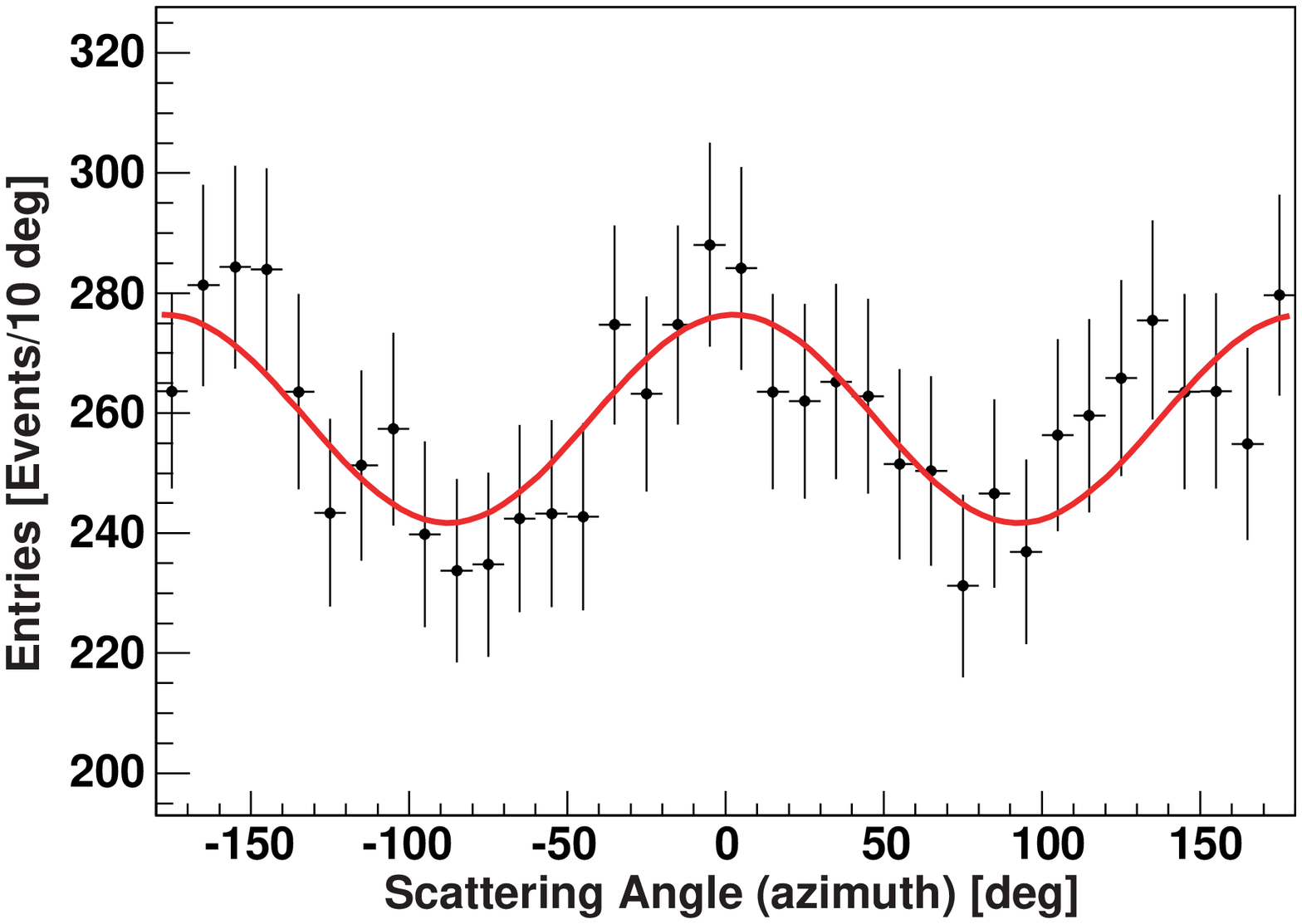}
   \end{tabular}
   \end{center}
\vspace*{-0.4cm}
   \caption[Azimuth Compton scattering angle distribution in the SGD from GRB 021206 (a) with 80\% linear polarization and (b) with 10\% linear polarization ($5\sigma$ detection limit).]
   { \label{fig:GRB-pol} Azimuth Compton scattering angle distribution in the SGD from GRB 021206 (a) with 80\% linear polarization and (b) with 10\% linear polarization ($5\sigma$ detection limit).}
   \end{figure}

Since the SGD is a narrow-FOV instrument, it requires a great luck to observe GRBs in its FOV.
In this study, we simulated a hypothetical instrument, one SGD unit without collimator, to evaluate its performance for very bright GRBs such as GRB 021206.
Figure~\ref{fig:GRB-pol} (a) shows the azimuth Compton scattering angle distribution in one SGD unit from GRB 021206 with 80\% linear polarization.
A fit yields $Q = 0.488\pm0.013$, corresponding to the polarization degree of $78\pm2$\%. 
($Q = 0.629\pm0.012$ for 100\% polarization.)
Figure~\ref{fig:GRB-pol} (b) shows the azimuth Compton scattering angle distribution in one SGD unit from GRB 021206 at 5$\sigma$ detection limit, {\it i.e.} 10\% polarization.
A fit yields $Q = 0.067\pm0.015$, corresponding to the polarization value of $11\pm2$\%.
These results imply that one SGD unit can detect the polarization of quite a few GRBs at a few 10\% level.
It should be noted that the semiconductor Compton telescope is quite suitable for the GRB observation because of its good timing properties such as the time resolution of $\sim$0.1~$\mu$s and the dead time of $\sim$10 $\mu$s .
Furthermore, the SMCT (a wide FOV instrument) will have 1\% or better 5$\sigma$ the polarization sensitivity for GRB 021206 and better than 10\% for majority of GRBs due to its large effective area.

\section{Experimental Validation}\label{sect:exp}

We have fabricated prototype detector elements for DSSD and CdTe pixel detectors and achieved excellent energy resolution; 1.3~keV (FWHM) for Si detectors and 1.6~keV (FWHM) for CdTe detectors at 60~keV.\cite{Tajima02,Nakazawa03,Fukazawa04,Tajima04}
Using these elements, we have assembled small prototype Compton telescopes that consist of one DSSD and two CdTe detectors.
We performed several measurements with the prototype Compton telescopes to validate Monte Carlo simulations and verified its predictions\cite{Tajima03,Mitani03}.
Some results are shown here.

\begin{figure}[bth]
   \begin{center}
   \includegraphics[height=5cm]{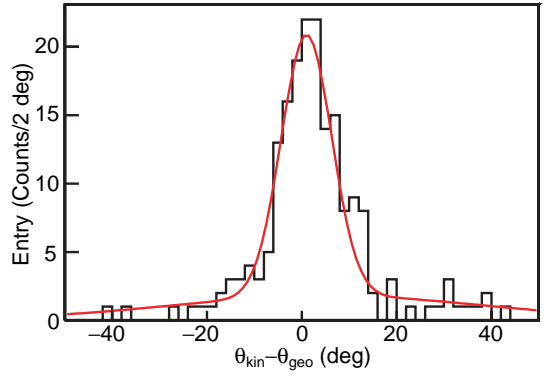}
   \end{center}
\vspace*{-0.4cm}
   \caption[Distribution of the difference between the Compton scattering angle calculated from the energy deposit and the geometrical angle obtained using a ${}^{57}$Co source.] 
   { \label{fig:Comp-angle} Distribution of the difference between the Compton scattering angle calculated from the energy deposit ($\theta_{\mathrm{kin}}$) and the geometrical angle ($\theta_{\mathrm{geo}}$) obtained using a ${}^{57}$Co source.}
   \end{figure}

Figure~\ref{fig:Comp-angle} shows the difference between the Compton scattering angle calculated from the energy deposit and the geometrical angle obtained using a ${}^{57}$Co source.
A fit to a sum of two Gaussian functions yields an angular resolution of 13.4\degree\ (FWHM), which is in good agreement with the MC result of 11.8\degree\cite{Tajima03}.
The dominant contributions to the angular resolution are estimated to be Doppler broadening (8.8\degree),  the energy resolution (6.0\degree) and the angular resolution of scattered photon (5.1\degree).
This analysis proves that Doppler broadening is the dominant component and the detector performance has already reached adequate level.
The latter two detector contributions can still be reduced by using a smaller strip pitch (DSSD had 800~\micron\ pitch instead of the design value, 400~\micron) and by improving the energy resolution. \footnote{In this test, the energy resolution was 2.5~keV while we have achieved better energy resolution of 1.3~keV in a different detector configuration.}

\begin{figure}[bth]
   \begin{center}
   \begin{tabular}{ll} 
   (a) Experimental result & (b) EGS4 simulation \\
   \includegraphics[height=5cm]{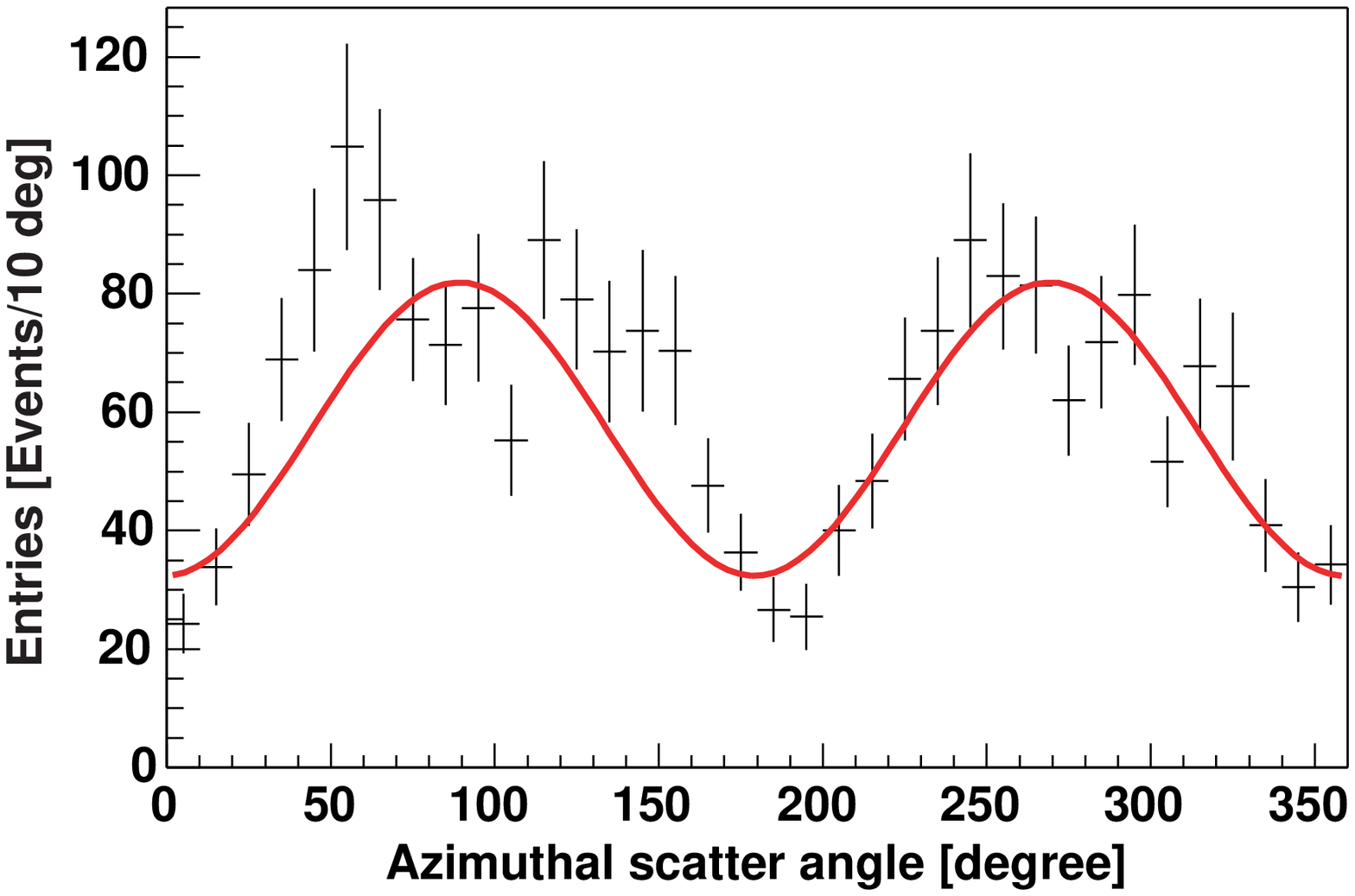} \hspace*{0.2cm} &
   \includegraphics[height=5cm]{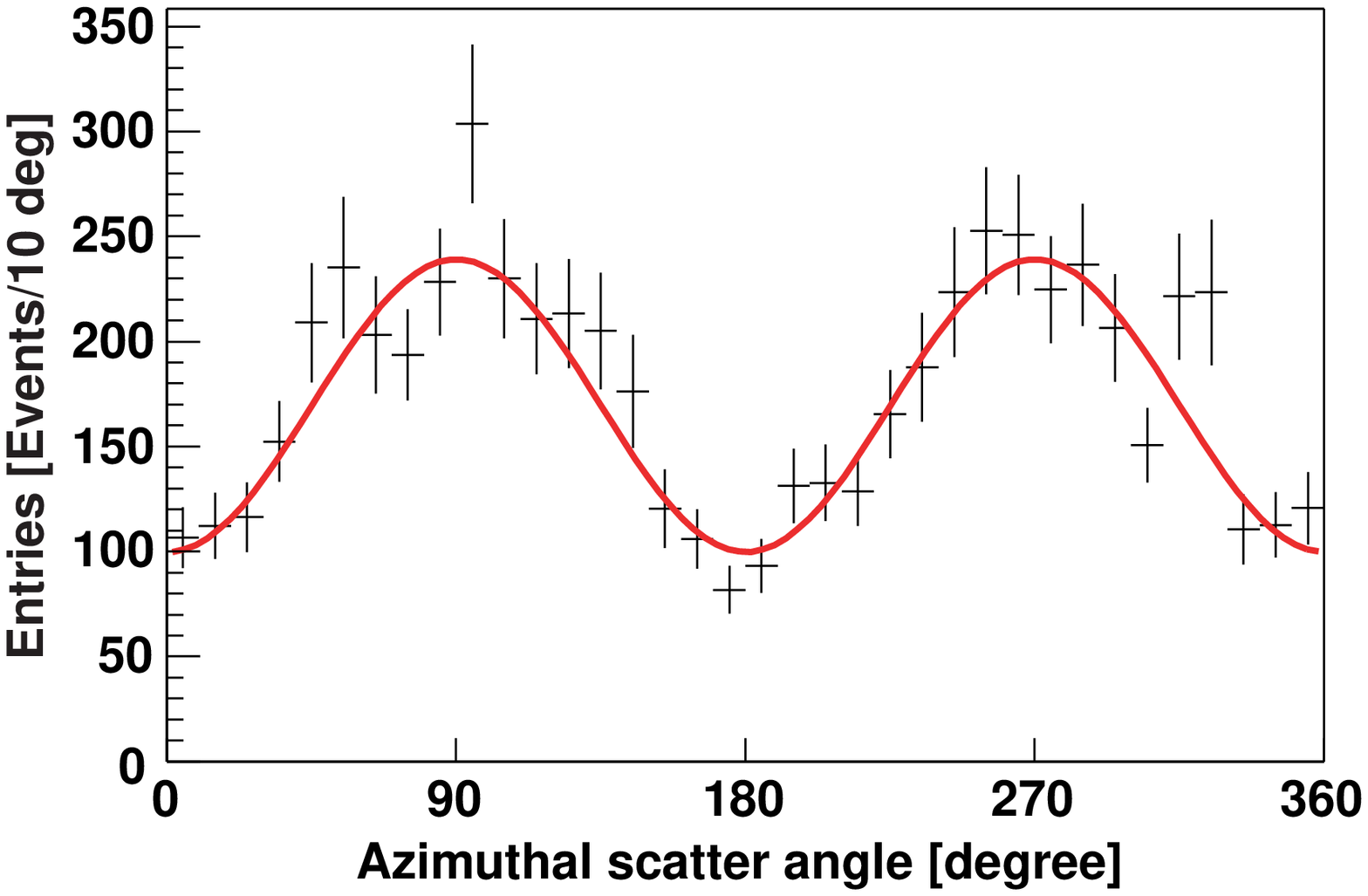}
   \end{tabular}
   \end{center}
\vspace*{-0.4cm}
   \caption[(a) Azimuth angle distribution of the Compton scattering obtained using a Compton telescope consists of a DSSD and CdTe pixel detectors with 177~keV polarized photon beam at SPring-8. (b) Corresponding distribution by EGS4 simulation.] 
   { \label{fig:sp8-phi} (a) Azimuth angle distribution of the Compton scattering obtained using a Compton camera consists of a DSSD and CdTe pixel detectors with 177~keV polarized photon beam at SPring-8. (b) Corresponding distribution by EGS4 simulation.}
   \end{figure}

In order to evaluate the polarization performance of the prototype Compton telescope,
we have carried out a beam test at SPring-8 photon factory in Japan.
Figure~\ref{fig:sp8-phi} (a) shows the azimuth angle distribution of the Compton scattering obtained with 177~keV photon beam with 100\% linear polarization.
Figure~\ref{fig:sp8-phi} (b) shows the corresponding distribution simulated by the EGS4.
A fit on the experimental measurement yields the modulation factor of $43\pm3$\% which is in a good agreement with the EGS4 result of $41\pm2$\%\cite{Mitani03}.
These results validate our implementation of the EGS4 simulation at 10\% level.


\section{Conclusions}
We have studied the SGD performance for the polarization measurements.
Measurements of the polarization degree and phase do not depend on these properties of incident photons.
No major systematic bias effects were seen at 1\% level.
Simulations with no polarization have revealed four bumps in the azimuth angle distribution corresponding to the square arrangement of the base module.
The effect is small and can be corrected.

We have demonstrated that 5$\sigma$ sensitivities of the SGD for bright sources such as Crab, Cygnus X-1 and Mkn 501 range from 1 to 2\%.
Such sensitivities is very likely to result in observations of hard X-ray polarization from these objects and will bring rich information to strengthen our understandings.
On the other hand, a lack of polarization will seriously undermine current models .
One SGD unit without collimator is shown to be quite sensitive to the polarization of bright GRBs and will provide a definitive answer on sizable polarization ($>$10\%) from GRBs.

\acknowledgments     

This work has been carried out under support of U.S. Department of Energy, contract DE-AC03-76SF00515,
Grantin-Aid by Ministry of Education, Culture, Sports, Science and Technology of Japan (12554006, 13304014),
and ``Ground-based Research Announcement for Space Utilization'' promoted by Japan Space Forum.

\bibliography{slac-pub9493}


\end{document}